\documentclass[journal ]{new-aiaa}
\usepackage[utf8]{inputenc}
\usepackage{textcomp}
\usepackage{subfigure}
\usepackage[normalem]{ulem}
\usepackage{graphicx}
\usepackage{amsmath}
\usepackage[version=4]{mhchem}
\usepackage{siunitx}
\usepackage{longtable,tabularx}
\title{Hypersonic Shock-Wave/Boundary-Layer Interaction on a Three-Dimensional Expansion-Compression Geometry}

\author{Anshuman Pandey \footnote{Assistant Professor, Department of Mechanical and Aerospace Engineering, AIAA Senior Member}}
\affil{University of South Florida, Tampa, FL, 33620, USA}
\author{Katya M. Casper \footnote{Distinguished Member of Technical Staff, Aerosciences Department, AIAA Associate Fellow}, Steven J. Beresh \footnote{Distinguished Member of Technical Staff, Aerosciences Department, AIAA Fellow}, Rajkumar Bhakta \footnote{Principal Technologist, Aerosciences Department, AIAA Member}, Marie E. De Zetter \footnote{Senior Technologist, Aerosciences Department} and Russell Spillers \footnote{Distinguished Technologist, Aerosciences Department}}
\affil{Sandia National Laboratories, Albuquerque, NM, 87185, USA}

\begin{document}

\maketitle

\begin{abstract}
This experimental work explores the flow field around a three-dimensional expansion-compression geometry on a slender cone at Mach 5 and 8 using high-frequency pressure sensors, high-framerate schlieren, temperature-sensitive paint, shear-stress measurements and oil-flow visualizations. The $7^\circ$ cone geometry has a hyperbolic slice acting as an expansion corner which is then followed by a $30^\circ$ finite-span compression ramp. The freestream Reynolds number was varied so that the boundary layer approaching the expansion corner was either laminar, transitional or turbulent. At laminar or early transitional conditions, the separation shock locks onto the expansion corner and the separation region encompasses most of the slice, with the separation shear layer flapping at a preferred frequency.  As Reynolds number is increased, the separation shock moves downstream onto the slice, the separation bubble shrinks, and the shear layer flapping frequency increases while its amplitude drops. In all cases, large-scale low-frequency breathing motions are observed.  The strong relaminarization across the expansion corner at Mach 8 prevents the shock/boundary-layer interaction from reaching truly turbulent conditions and fundamentally changes its behavior on this non-canonical geometry.
\end{abstract}

\section*{Nomenclature}
{\renewcommand\arraystretch{1.0}
\noindent\begin{longtable*}{@{}l @{\quad=\quad} l@{}}
$P_0$  & stagnation pressure \\
$T_0$  & stagnation temperature \\
$L$& length of the separation bubble in the plane of symmetry \\
$L_{slice}$& length of the slice between the cone and the start of the ramp in the plane of symmetry $=94.11$ mm \\
$x,y,z$ & coordinates centered at the origin \\
$\bar{x},\bar{y},\bar{z}$ & coordinated normalized by $L_{slice}$ \\
$\delta$   & boundary-layer thickness obtained from mean schlieren image \\
$U_{e}$ & cone-edge velocity obtained from Taylor-Maccoll equations \\
$f$ & frequency \\
$c_{h}$ & heat-flux coefficient \\
$P_{ce}$   & cone-edge pressure \\
$P_{pm}$   & pressure after a 2D inviscid expansion \\
$P_{is}$   & pressure after a 2D inviscid shock \\
$P$   & mean pressure \\
$h$  & height of the shear layer \\
$St$  & Strouhal number based on $h$ and $U_e$ \\
$St_{\delta}$  & Strouhal number based on $\delta$ and $U_e$ \\
$St_{L}$  & Strouhal number based on $L$ and $U_e$ \\
$\rho$ & density\\
\end{longtable*}}

\section{Introduction}
\lettrine{T}{he} aerodynamic environment surrounding a hypersonic vehicle is characterized by a multitude of intricate flow phenomena. Flow deflection into the freestream, such as that obtained by a compression corner, results in shock waves that can interact with the upstream boundary layer, amplify disturbances, and cause it to separate \citep{Gadd1954, Chapman1958}. On the other hand, flow deflection away from the freestream, such as that by an expansion corner, results in expansion fans and attenuation of disturbances in the boundary layer \citep{Sternberg1954}. The accurate prediction of this boundary-layer response, in conjunction with the phenomenon of shock-wave/boundary-layer interaction (SBLI), has been recognized as a pivotal scientific impediment to the advancement of hypersonic vehicle technology \citep{Leyva2017}.

   The SBLI phenomenon represents a cornerstone of high-speed aerodynamics, fundamentally influencing the performance, stability, and structural integrity of supersonic and hypersonic vehicles. These interactions can lead to a cascade of responses including significant increases in local pressure and heat transfer, boundary-layer separation, and the generation of severe unsteady aerodynamic and heating loads \citep{Green1970, Delery1985, Knight2003, Gaitonde2023ARFM}. As the name suggests, SBLI characteristics are dependent upon the nature of the incoming boundary layer and the strength of the adverse pressure gradient imposed by the shock wave. A laminar boundary layer is highly susceptible to separation whereas a turbulent boundary layer has increased momentum in the near-wall region that enables it to withstand a stronger adverse pressure gradient \citep{Arnal2004}. Separation of a fully turbulent boundary layer is highly dependent on its state as captured by the shape parameter or other measure of the near-wall momentum.

\begin{figure} [hbt!]
		\centering
		{\includegraphics[trim={120 0 150 0}, clip, width=1\textwidth]{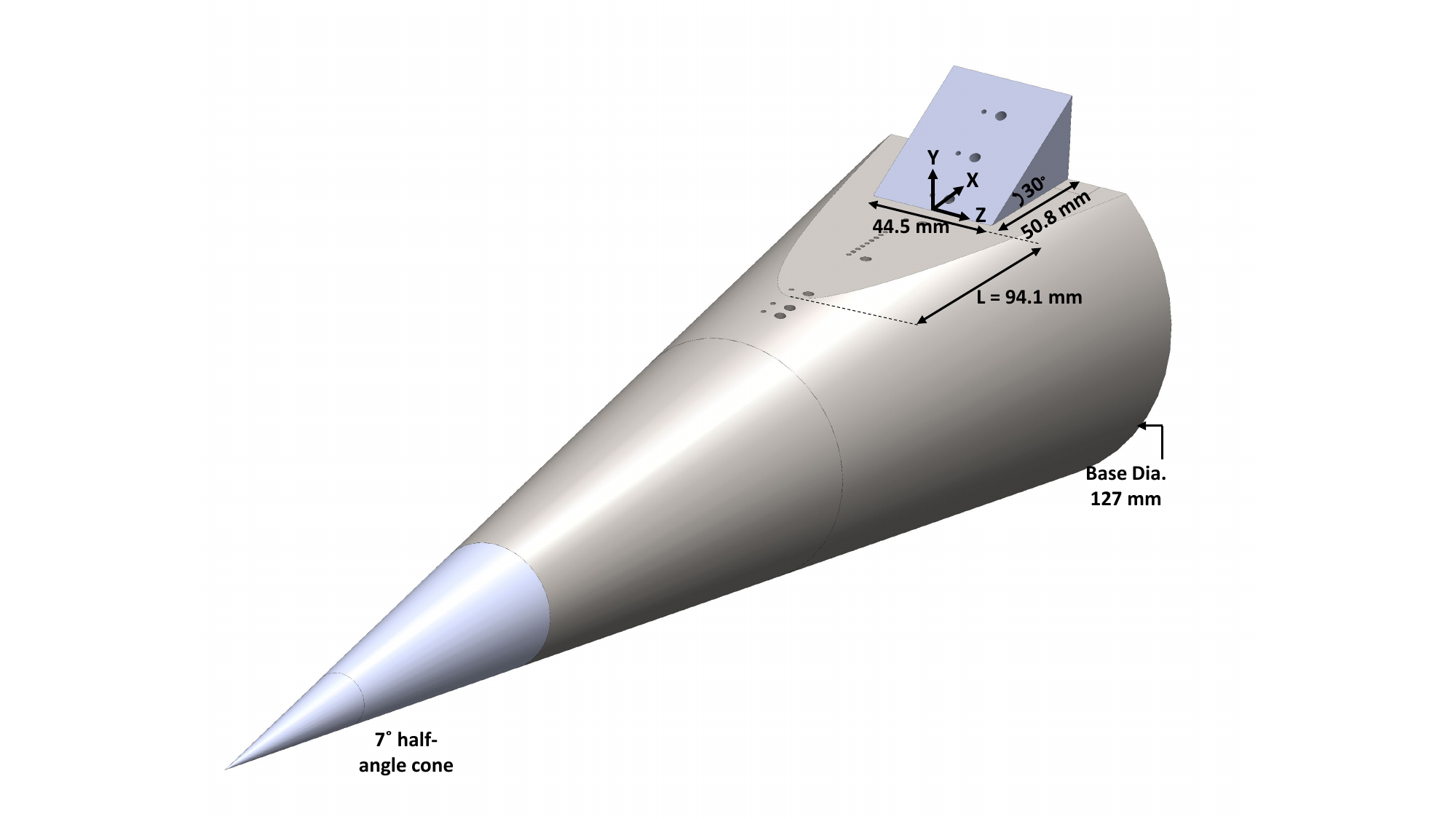}}
		\caption{Cone-slice-ramp geometry with coordinate axes used in this work.}
		\label{fig:FigGeometry}
	\end{figure}

    Interaction of a high-speed boundary layer with a two-dimensional compression ramp is a canonical SBLI geometry that has received considerable attention \cite{Delery1985, Knight2003, Clemens2014}. Extensive investigations also have employed two-dimensional canonical geometries to elucidate the effects on a boundary layer due to expansion features, but these ordinarily have remained separate from those studying compression corners \citep{Smits2006}. However, as highlighted by Dolling \cite{Dolling2001}, new studies on complex geometries are essential to transcend the limitations of canonical investigations and bridge the existing knowledge gap towards flight applications. Such geometries often possess both expansion and compression features. Accordingly, the present study centers on a non-canonical geometry, depicted in figure \ref{fig:FigGeometry}, inspired by previous hypersonic research concepts e.g., \cite{Oberkampf1992, Walker1992, Massobrio2007}. In these shapes, an upstream expansion alters the state of the boundary layer that travels downstream to generate a SBLI, and hence the nature of the SBLI becomes dependent on this altered boundary-layer state.

    In the current work, the shock wave is generated by a finite-width ramp of fixed angle that is located on a slice carved through a sharp cone.  The junction of the conical frustrum with the slice creates an expansion corner of hyperbolic shape.  Thus the combination of upstream expansion and downstream compression becomes highly three-dimensional. The state of the boundary layer is varied in these experiments by changing the freestream Reynolds number $(Re)$ to enable the study of laminar, transitional, and turbulent interactions on this geometry.  Furthermore, experiments have been conducted to test these three flow regimes at freestream Mach numbers $(M)$ of both 5 and 8.
    
   From the large body of canonical research, it is known that a laminar boundary layer is most prone to separation, and consequently laminar SBLI’s exhibit large separation regions e.g., \cite{Lees1964, Miller1964, Katzer1989}. This extensive separation causes significant deviations in surface-pressure distributions compared to the abrupt pressure jump predicted by the inviscid (unseparated) oblique-shock theory. A distinct decrease in both skin friction and heat transfer is observed upon boundary layer separation that increases readily near reattachment. For large ramp angles, secondary separations can form within the primary separation zone that can further alter these surface distributions \citep{Korolev2002, Gai2019}. Even a nominally two-dimensional laminar interaction develops three-dimensionality and unsteadiness through the growth of convective and global instabilities in the separation bubble. For a constant ramp angle, variations in $M$ and $Re$ lead to alterations in the separation bubble size \citep{Chapman1958, Miller1964}. Furthermore, spanwise-curvature of axisymmetric geometries also reduce the size of the separation region due to the three-dimensional relieving effect \citep{Huang1983}.

    With increasing freestream $Re$ and the strength of the disturbance environment, the aforementioned instabilities of the laminar flow cause eventual breakdown to turbulence. As the transition location moves progressively upstream, the breakdown can begin to occur in the separated shear layer or in the incoming boundary-layer itself. Several studies have focused on one or both of these regimes \cite{Simeonides1995, Benay2006, Sandham2014, Murphree2021, Butler2022} and it is now known that the scale of the transitional SBLI demonstrates extreme sensitivity to the transition location. The first regime, where a laminar boundary layer separates but begins to breakdown at reattachment, is frequently associated with the highest recorded surface heating rates, often exceeding those in fully laminar or turbulent interactions, due to the intermittent and highly energetic nature of the transitional structures \citep{Simeonides1995}. Computational studies \citep{Sansica2016, Saidi2025} have reported that the broadband unsteadiness at reattachment can travel upstream and selectively feed low-frequency unsteadiness of an otherwise laminar separation. In the second regime, where separation occurs for a transitional boundary layer, passage of turbulent spots from the upstream boundary layer can dramatically alter the separation size and the resulting surface-pressure distribution \citep{Krishnan2007, Estruch-Samper2022}. Turbulent spots induce a temporarily fuller boundary layer that is less susceptible to separation \citep{Beresh2002} thereby intermittently collapsing the separation bubble. 

    Turbulent SBLI arises when the incoming boundary layer has transitioned fully to turbulence upstream of the interaction region. Numerous studies have focused on characterizing the scale of two- and three-dimensional turbulent SBLI \cite{Delery1985, Souverein2013, Gaitonde2023ARFM}. In the turbulent case, the separation region is smaller than for a laminar incoming boundary layer. The scale of the SBLI is dependent on the geometry, strength of the interaction (ramp angle or more generally, the pressure ratio across the shock that induces flow separation), wall temperature, Mach number, Reynolds number, and boundary-layer shape parameters. In hypersonic turbulent SBLI’s, the higher compressibility and energy content can influence the turbulence structure and its response to the adverse pressure gradient imposed by the shock, potentially leading to different scaling laws for separation size and unsteadiness compared to supersonic cases \citep{Smits2006}.

   In comparison to the canonical SBLI reviewed above, the geometry considered here (figure \ref{fig:FigGeometry}) includes an expansion upstream of the ramp which adds additional complexity. The expansion through its combined effect of streamline curvature, favorable pressure gradient, and dilatation (reduction in density) acts to stabilize the boundary layer. In laminar cases, the expansion corner leads to attenuation of second-mode instabilities developing on the cone \citep{Butler2021ExiF, Chuvakhov2021, Pandey2024} whereas in turbulent cases, there is an apparent reduction in the boundary-layer fluctuations \citep{Dussauge1987, Tichenor2013, Teramoto2017, Sun2017, Nicholson2024, Pandey2024}. 
   
   The seminal study by Zheltovodov et al. \cite{Zheltovodov1993} demonstrated that tandem expansion-compression SBLI’s can lead to larger separation sizes when compared to the canonical compression-corner SBLI \citep{Huo2022}. Computational simulations \citep{Knight2003, Fang2015, Xie2022} on this configuration have revealed that reduced mixing in the expanded boundary layer causes attenuation of near-wall fluctuations and velocity-profile fullness that result in increased susceptibility to separation. Furthermore,  the separation location can `lock-in' to the expansion corner \citep{Guo2024} due to reduced pressures downstream of expansion. In hypersonic flow, the cone-cylinder-flare (CCF) geometry has been used, however the focus has been to study SBLI on an axisymmetric geometry \citep{Schaefer1962, Gray1967} rather than to systematically explore the expansion-corner effects on the downstream SBLI. Recent work \cite{Paredes2022, Benitez2023, Benitez2025} on CCF has focused on the laminar or early-transition regime and several similarities have been noted with instabilities in hypersonic shear layers on other axisymmetric bodies such as a cone-flare \citep{Butler2022, Davami2025} or a hollow cylinder flare \citep{Lugrin2022}.
   
   The overarching goal of the current study is to understand how decades of research into two-dimensional SBLI translates to the cone-slice-ramp geometry shown in figure \ref{fig:FigGeometry}. The current work compiles several experimental campaigns in the Sandia Hypersonic Wind Tunnel (HWT) to characterize the hypersonic flow around this non-canonical geometry across varying Reynolds and Mach numbers and supplements other published studies by the same authors \citep{Pandey2023, Pandey2024}. These studies complement the laminar SBLI work carried out in the 1990s in the HWT by Oberkampf et al. \cite{Oberkampf1992} and recently in other quiet \citep{McKiernan2021, Nicotra2024} and high-enthalpy wind-tunnels \citep{Stramenga2025, Jans2025}. These laminar experiments have been used for validation of laminar simulations \citep{Walker1992, Thome2018, Sadagopan2023}, and the current dataset would assist development of predictive tools on such geometries in transitional and turbulent regimes. 
   
    First, the phenomenon of boundary-layer relaminarization due to the expansion corner is established across the flow regimes. A remarkable effect of relaminarization is the enhanced susceptibility to separation, which is identified in tests at Mach 5 and Mach 8. Next, the mean flow characteristics are presented to show the size of the separation region and its dependence on the relaminarization phenomenon and the flow regime. Lastly, the unsteady characteristics of the interaction are quantified to show both narrow-band and broad-band unsteadiness that evolves with $Re$ and $M$. Together, these features highlight the greater complexity of the SBLI on this non-canonical geometry as compared to the simpler geometries that fill the literature.

\section{Experimental Setup}
	\label{section:ExperimentalSetup}

\subsection{Sandia Hypersonic Wind Tunnel and Test Conditions}
	The Sandia Hypersonic Wind Tunnel (HWT) is a conventional blowdown-to-vacuum facility with an interchangeable system of contoured nozzles and heater sections for the selection of a desired Mach number in the test section \citep{Beresh2015}. In this work, the Mach 5 and 8 systems have been used and their operational details are summarized in table \ref{tab:table1}. The noise levels were quantified as the RMS Pitot pressure between 0 to 50 kHz over the mean Pitot pressure and were within the reported range for all $Re$ tested \citep{Casper2016}. 

\subsection{Model and Sensors}
	
	The wind-tunnel model shown in figure \ref{fig:FigGeometry} is a $7^{\circ}$ half-angle slender cone with a base diameter of 0.127 m. The nosetip was nominally sharp with a radius less than 0.05 mm. The cone has a hyperbolic cut at an offset of 45.71 mm from its axis. The axial length of the resulting horizontal slice is 144.91 mm, and at its aft-end the slice has a provision to mount 50.8 mm long and 44.45-mm wide ramps of different deflection angles. In this work, a flat plate ($0^{\circ}$ ramp) and a $30^{\circ}$ ramp have been used for the majority of the discussion while data from other ramp angles of $10^{\circ}, 20^{\circ}$ and $40^{\circ}$ are introduced as needed. The origin was defined at the slice-ramp corner and at midspan of the ramp with $(x,y,z)$ coordinates pointing in the axial, slice-normal and spanwise directions, respectively. In this work, the coordinates have been non-dimensionalized using the distance of the cone-slice expansion corner to the base of the compression corner, i.e., $(\bar{x},\bar{y},\bar{z})$ = $(x,y,z)/L_{slice}$ where $L_{slice} = 94.11$ mm. This model was mounted using a sting at the base of the cone and all testing was done at zero angle of attack. The experiments were conducted over several different HWT campaigns with compatible measurement/visualization techniques.

\begin{table}[hbt!]
\caption{\label{tab:table1} Sandia Hypersonic Wind Tunnel (HWT) operational capabilities}
\centering
\begin{tabular}{lcccccc}
\hline
Mach    &  Medium  &  	$P_0$ (kPa)	&  	$T_0$ (K)	&  	$Re \times 10^6$ /m &  Noise level ($\%$)\\\hline
5       &  Air     &	345--1380 	& 	330--890 	&   3.3--26             &  1--2    \\  
				8  		&  Nitrogen&	1720--6890 	& 	500--890   	& 	3.3--20 	        &  3--5    \\
\hline
\end{tabular}
\end{table}

The sensors used in this work have been placed near the plane of symmetry of this three-dimensional geometry. Two rows of sensor holes, each offset from the plane of symmetry by 3.20 mm, were used. One row has 1.76 mm diameter holes for mounting Kulite pressure sensors and the other row has 3.81 mm diameter holes for mounting PCB pressure sensors, Schmidt-Boelter heat-flux sensors and Ahmic shear-stress sensors. Different sensors may occupy the same hole during different wind tunnel entries. The holes on the ramp are at the same distance from the origin irrespective of the ramp angle. The hole locations and the sensors used in the experiments have been summarized in table \ref{tab:tablesensor}. A top-down view of the sensor configuration is available in \cite{PandeyST2022}. The sensors were mounted flush with the model surface and the sensor cables routed through the sting to the data-acquisition system outside the tunnel. The model base plate had holes that allowed the internal cavity to equilibrate with the low pressures in the HWT. A Kulite ETL-79 was placed in the model cavity to provide accurate measurements of the pre-run vacuum pressures and aided in correcting the zero bias of the Kulite sensors on the model.
	
	The Kulite row had a combination of Kulite XCQ-062-30As or XCE-062-15As for measuring surface-pressure fluctuations. These A-screen sensors have flat response up to 40 kHz and compare favorably with the higher-frequency PCB sensors up until 60 kHz; see figure 3 in \cite{Beresh2010} for a comparison in a Mach 2 turbulent boundary layer. The sensors have a reported uncertainty of only 0.5$\%$ of full-scale output. Precision Filters 28000 analog signal-conditioning system with a 28144 module was used to provide the excitation voltage to the sensors and also to low-pass filter (anti-alias) the signals. The digitization was carried out using an NI-PXI system with NI-6133 modules. The signals were low-pass filtered at 80 kHz and sampled at 500 kHz. 
	
	In this work, PCB132B38 sensors that have a frequency bandwidth of 11-1000 kHz have been used that allow measurements of high-frequency pressure fluctuations in laminar \citep{Fujii2006} and turbulent boundary layers \citep{Beresh2011}. These sensors have demonstrated success in previous HWT experiments on a straight-cone \citep{Casper2016} and extend the measurements of the lower-frequency Kulite sensors to capture the second-mode waves and its higher harmonics. The sensor has an external diameter of 3.175 mm and a rubber sleeve was used to help isolate it from the rest of the structure; the actual sensing-element diameter is 0.81 mm \citep{Ort2019}. The excitation was provided by a PCB 482A22 signal conditioning system and the anti-alias low-pass filtering was provided by a Precision Filters 28612 module. The digitization was carried out using an NI-PXI system with NI-6396 modules. The signals were low-pass filtered at 2 MHz and sampled at 5 MHz. In measurements of high-frequency, small-scale disturbances, spatial averaging effects due to the finite sensor size can cause unwanted attenuation of fluctuations \citep{Beresh2011, Huang2024}. Where applicable, the \cite{Corcos1963} correction has been performed to correct for this attenuation.	
	
	Schmidt-Boelter gauges (Medtherm Corporation 8-1-0.25-45-20835EBS) were used to measure the heat flux on the slice and the ramp. The gauges house a thermopile in a 2.79 mm diameter cavity with a thermocouple on each of its two ends \citep{Sullivan2012}. The sensor has an external diameter of 3.175 mm and the same mounting procedure as the PCB was used. The sensor has a reported uncertainty of 3$\%$. The signals were low-pass filtered at 1 kHz and were sampled at 20 kHz. 
	
	Shear-stress measurements were carried out using novel miniature sensors (AH-125-02) developed by Ahmic Aerospace, LLC. The sensing element was a floating head that was connected to the base of the housing through a flexure connected to strain gauges. After calibration of the strain induced on the flexure, measurement of shear stress on the floating head can be made \citep{Meritt2016, Meritt2017}. The sensing element and the external housing are 2.41 mm and 3.175 mm in diameter, respectively. The sensor can measure shear stress up to 150 Pa with an uncertainty of 0.75 Pa at a rate of about 250 Hz. The signals were low-pass filtered at 1 kHz and sampled at 20 kHz. 

\begin{table}[hbt!]
\caption{\label{tab:tablesensor} Instrumentation locations (distances from origin) on the model with lines demarcating the sensors on the cone, slice, and ramp. $\bulletSS$ denotes corresponding measurements were made.}
\centering
\begin{tabular}{lcccccc}
\hline
Location $x$ (mm)  			&  	PCB   			&  	Kulite 		&  	Schmidt-Boelter &  Shear-stress\\\hline
-106.6    				& 	$\bulletSS$ 	& 	$\bulletSS$ 	&   				& 	 				\\
				-100.3   				& 	$\bulletSS$ 	& 	$\bulletSS$   	& 					&	$\bulletSS$ 	\\
				\hline
				-88.0     				& 	$\bulletSS$ 	& 	$\bulletSS$   	&   				& 	$\bulletSS$ 	\\
				-49.9     				& 	$\bulletSS$ 	& 	$\bulletSS$   	& 	$\bulletSS$ 	& 	 			 	\\
				-46.85     				& 					& 	$\bulletSS$   	&   				& 				 	\\
				-43.8     				& 					& 	$\bulletSS$   	&   				& 				 	\\
				-40.75     				& 					& 	$\bulletSS$   	&   				& 				 	\\
				-37.7     				& 					& 	$\bulletSS$   	&   				& 				 	\\
				-34.65     				& 					& 	$\bulletSS$   	&   				& 				 	\\
				-31.6     				& 					& 	$\bulletSS$   	&   				& 				 	\\
				-28.55     				& 					& 	$\bulletSS$   	&   				& 				 	\\
				-25.5     				& 					& 	$\bulletSS$   	&   				& 				 	\\
				-11.8     				& 	$\bulletSS$ 	& 	$\bulletSS$   	&   				& 	$\bulletSS$	 	\\
				\hline
				6.1     				& 	$\bulletSS$ 	& 	$\bulletSS$   	& $\bulletSS$  		& 				 	\\
				24.9     				& 	$\bulletSS$ 	& 	$\bulletSS$   	& 	$\bulletSS$ 	& 	$\bulletSS$	 	\\
				37.9 ($0^{\circ}$)     	& 	$\bulletSS$ 	& 	$\bulletSS$   	&   				& 				 	\\
				43.7 ($30^{\circ}$)    	& 	$\bulletSS$ 	& 	$\bulletSS$   	& $\bulletSS$  		& 				 	\\
\hline
\end{tabular}
\end{table}

	\subsection{High-framerate Schlieren}
	\label{sec:Schlieren_FLDI}
	
	The high-framerate schlieren system incorporated a high repetition-rate pulsed laser (Cavilux Smart) and a high-frame-rate camera (Phantom v2512 or Phantom TMX7510). The laser generated pulses at a repetition rate of 100 kHz where the pulse duration was 10 ns each. The z-type schlieren configuration used two 450.8-mm diameter, 2.75-m focal length, spherical mirrors for light-collimation and a knife edge oriented horizontally to resolve vertical gradients in density \citep{Settles2001}. The schlieren system provided a spanwise-integrated visualization of the boundary layer and SBLI. The typical camera configuration used a 640 x 208 pixel area to capture the whole slice-ramp region with a resolution of 2.56 pixels/mm.

	\subsection{Oil-flow Visualization \& Temperature-Sensitive Paint}
	
	To capture the three-dimensional aspects of the flow field, full-field surface visualization was carried out using oil flow and temperature-sensitive paint (TSP). Oil-flow visualization (OFV) provided surface streamlines on the geometry which were useful in verifying flow symmetry as well as in identifying separation lines. A thin coating of low-viscosity fluorescent liquid (Zyglo ZL-15) was sprayed on the model prior to the run. Over the course of the run, the oil coating flowed and aligned with the surface streamlines. It was illuminated using UV lights and the slice-ramp region was captured by a LaVision sCMOS camera at a resolution of 15.1 pixels/mm. OFV data was unavailable on the reattachment ramp due to divergence of streaklines and higher temperatures that caused evaporation of oil.
	
	TSP was used to visualize the heating pattern on the model and provided an additional diagnostic for the extent of the SBLI separation region, especially on the ramp where OFV was unusable as mentioned above. The TSP formulation used was Ru(bpy) in a clearcoat \citep{Liu2005}. This formulations essentially provides a steady-state measurement compared to characteristic flow times. The paint was excited with 460-nm water-cooled lights. Images were acquired with a LaVision sCMOS camera at a resolution of 6.8 pixels/mm. A 550 nm high-pass filter was used to remove the excitation light and capture only the paint emission. The temperature measured by TSP was converted to heat flux using an in-situ scaling method used in previous studies \citep{Liu2013, Juliano2015}. A scale factor was computed by using a linear fit between the heat-flux obtained from a reference Schmidt-Boelter gauge, located on the slice ($x$ = -49.9 mm), and the temperature obtained from TSP near the gauge. Under the assumption that the paint characteristics were similar and the heating effects were linear, this scale factor was then applied to the whole TSP image to estimate the full-field heat flux. Although the accuracy of this scaling method is limited further away from the reference gauge location \citep{Juliano2015}, it does not affect the inferences made in this work where the TSP data was predominantly used for visualization of the separation bubble and its features.

\section{Boundary layer development due to the expansion slice}
	\label{section:ExpansionResults}

The evolution of the boundary layer from the cone to the slice is investigated to understand the influence of three-dimensional expansion on boundary-layer characteristics. This section primarily focuses on the Mach 5 dataset, with comparisons drawn from Mach 8 data previously detailed in \cite{Pandey2024}.

\begin{figure} 
		\centering
		\subfigure[]
		{\includegraphics[trim={150 230 140 250}, clip, width=0.45\textwidth]{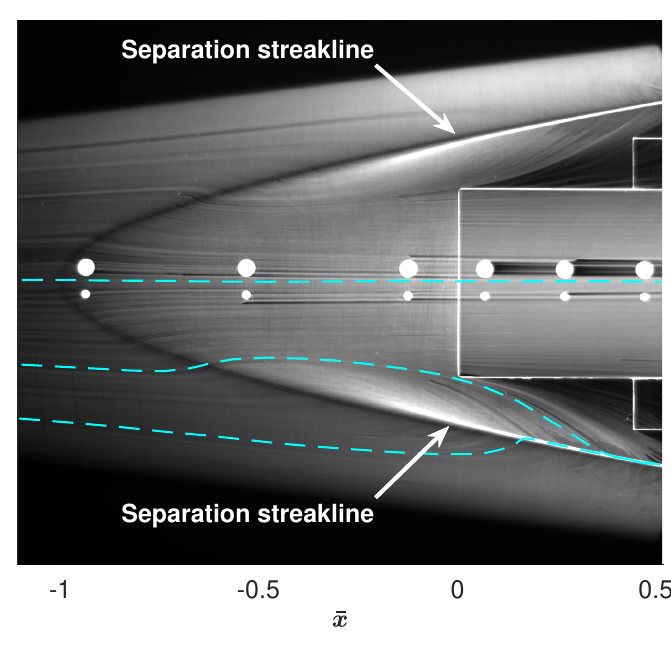} \label{fig:OFV_0deg_TuM5}}
		\subfigure[]
		{\includegraphics[trim={150 230 140 250}, clip, width=0.45\textwidth]{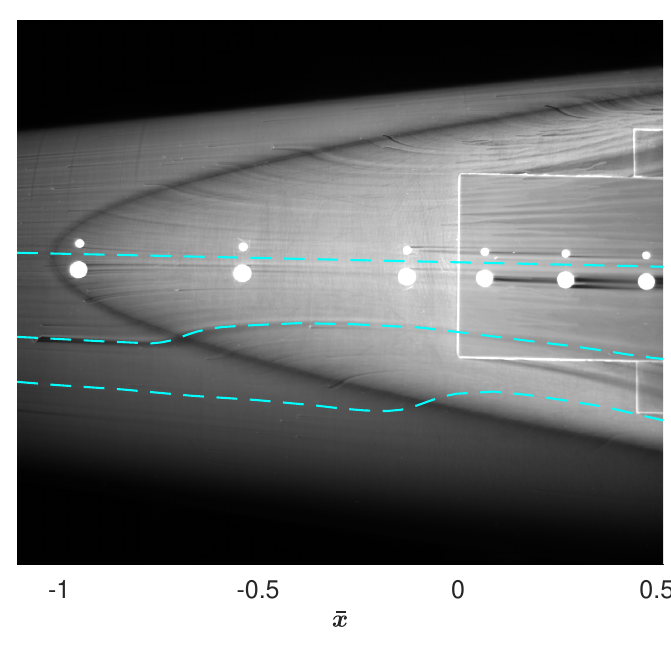} \label{fig:OFV_0deg_TuM8}}
		\caption{Oil flow visualization for the cone-slice geometry ($0^{\circ}$ ramp) with a turbulent inflow (a) Mach 5, $Re = 16.0 \times 10^6$ /m  (b) Mach 8, $Re = 16.2 \times 10^6$ /m. Cyan colored dashed curves highlight a few streaklines.}
		\label{fig:OFV0deg}
	\end{figure}

    Figure \ref{fig:OFV0deg} presents instantaneous surface oil-flow visualizations for turbulent cases at (a) Mach 5 and (b) Mach 8, both at comparable freestream Reynolds numbers. While oil-flow streaklines exhibited negligible variation with Reynolds number (not shown), notable differences emerged between the two Mach numbers. The axial slice on the cone generally induces a low-pressure region, generating azimuthal flow. Upstream of $\bar{x} = -0.5$, spanwise flow is observed climbing onto the slice for both Mach numbers. Further downstream, a line of convergence, indicative of flow separation, appears on both shoulders of the slice in the Mach 5 case, commencing at approximately $\bar{x} = -0.15$. This spanwise separation at the slice shoulders was also observed in the computational study by \cite{Vogel2019}, which utilized a considerably larger slice geometry at Mach 6. The spanwise flow lifts off the slice shoulder and curls into streamwise circulation. Conversely, in the Mach 8 case shown in figure \ref{fig:OFV_0deg_TuM8}, while spanwise-outward curving of slice streaklines is observed, flow separation at the shoulders is absent. Streaklines from the cone continue to climb onto the slice until the end of the geometry. 

\begin{figure} [hbt!]
		\centering
		\subfigure[]
		{\includegraphics[trim={100 335 120 355}, clip, width=0.75\textwidth]{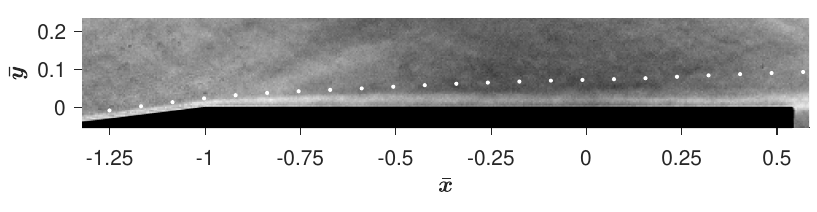} \label{fig:Sch0deg_Lam}} 
		\subfigure[]
		{\includegraphics[trim={100 335 120 355}, clip, width=0.75\textwidth]{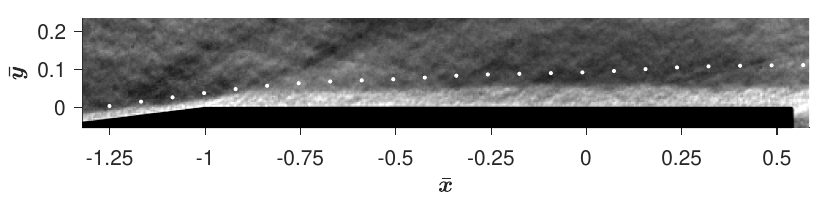} \label{fig:Sch0deg_Turb}} 
		\caption{Comparison of boundary-layer expansion for the cone-slice geometry ($0^{\circ}$ ramp) at Mach 5 and Mach 8. (a) Laminar case: Instantaneous image at Mach 5, $Re = 3.6 \times 10^6$ /m with white dots marking the boundary layer edge for Mach 8, $Re = 4.7 \times 10^6$ /m  (b) Turbulent case: Instantaneous image at Mach 5, $Re = 16.4 \times 10^6$ /m with white dots marking the boundary layer edge for Mach 8, $Re = 16.2 \times 10^6$ /m}
		\label{fig:Sch0deg}
	\end{figure}

Boundary-layer thickness, a critical factor in flow development, varies significantly with Mach number. Figure \ref{fig:Sch0deg} presents instantaneous schlieren images for Mach 5 under both laminar and turbulent conditions, illustrating the boundary layer as a lighter region along the geometry. Overlaid white dots indicate the boundary layer edges extracted from mean schlieren images for Mach 8. As found in \cite{Pandey2024}, the Mach 8 boundary layer is thicker pre-expansion. This disparity is amplified post-expansion due to the increased density change at higher Mach numbers, while the edge velocity remains relatively constant. Consequently, the principle of conservation of mass necessitates a more pronounced increase in boundary layer thickness with increasing Mach number, as evidenced in figure \ref{fig:Sch0deg}.
    
\begin{figure} [hbt!]
		\centering
		\subfigure[]
		{\includegraphics[trim={155 265 180 275}, clip, width=0.491\textwidth]{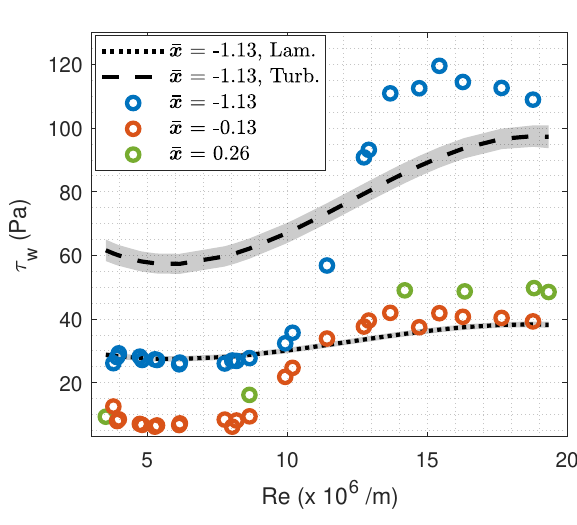} \label{fig:SS_0deg_allRe_M5}}
		\subfigure[]
		{\includegraphics[trim={155 265 180 275}, clip, width=0.491\textwidth]{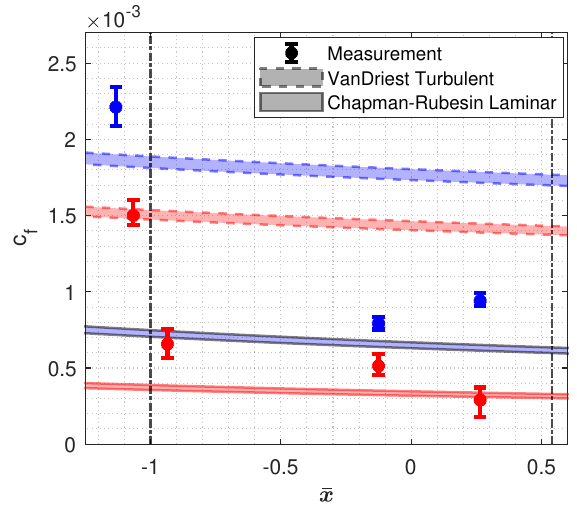} \label{fig:Cf_0deg_highRe}}
		\caption{Wall shear-stress measurements for the cone-slice geometry ($0^{\circ}$ ramp) (a) Mach 5, compared with turbulent and laminar correlations for a straight cone from \cite{VanDriest1956} and \cite{Chapman1949}, respectively (b) Comparison of high $Re$ cases across Mach 5 (blue) and 8 (red). Mach 5 cases are $Re > 16 \times 10^6$ /m and Mach 8 cases are $Re > 14 \times 10^6$ /m. Measurements indicated with uncertainty bars represent variation across different runs. Correlations for a straight cone are also shown.}
		\label{fig:SS0deg}
	\end{figure}

    Shear-stress measurements were conducted along a cone-slice geometry for Mach 5 across a range of Reynolds numbers, complementing existing Mach 8 data from \cite{Pandey2024}. Figure \ref{fig:SS_0deg_allRe_M5} presents these measurements at three locations: one on the upstream cone and two along the slice. These values are compared against shear-stress correlations for laminar \cite{Chapman1949} and turbulent \cite{VanDriest1956} boundary layers on a straight cone. Along the slice, shear-stress values are lower than on the cone. This reduction is attributed to decreased density and diminished boundary-layer momentum across the expansion. In the laminar regime, the shear-stress values at a particular location remain relatively constant with $Re$. As $Re$ increases and the boundary layer gradually transitions, shear stress values consequently rise. The transition front crosses the downstream sensor ($\bar{x} = 0.26$) at $Re \approx 8.0 \times 10^6$ /m, subsequently reaching upstream of the slice at $Re \approx 10.0 \times 10^6$ /m. Further increases in freestream $Re$ reveal a clear transition from laminar to turbulent correlation values at the cone sensor ($\bar{x} = -1.13$), marked by a typical transitional overshoot region. High $Re$ cases ($Re > 16 \times 10^6$ /m for Mach 5 and $Re > 14 \times 10^6$ /m for Mach 8) have been non-dimensionalized into a skin-friction coefficient using cone edge quantities and compared in figure \ref{fig:Cf_0deg_highRe}. As discussed in \cite{Pandey2024}, Mach 8 data (red marker) exhibit a continuous decrease in skin friction until the geometry's end due to relaminarization effects from the expansion. In contrast, Mach 5 data (blue marker) show a slight increase towards the end of the geometry.
    
\begin{figure} [hbt!]
		\centering
		\subfigure[]
		{\includegraphics[trim={130 240 140 255}, clip, width=0.49\textwidth]{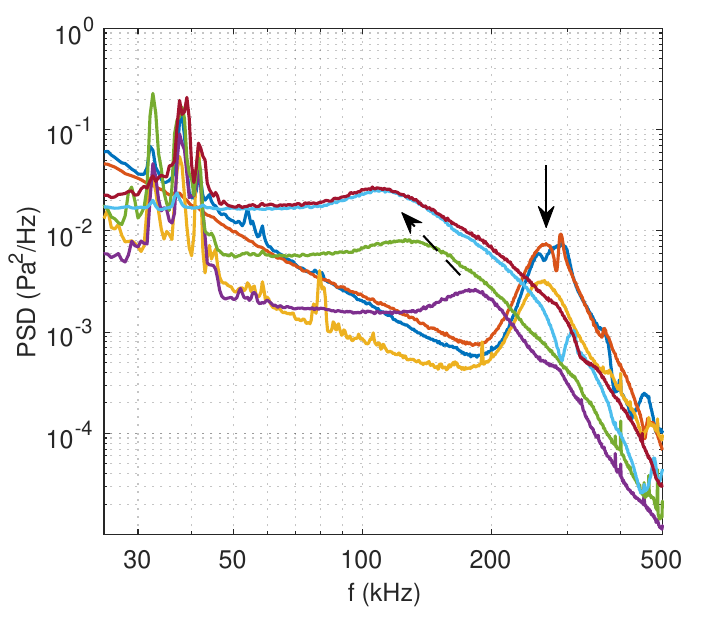} \label{fig:PSD_PCB_0_deg_Lam}}
		\subfigure[]
		{\includegraphics[trim={130 240 140 255}, clip, width=0.49\textwidth]{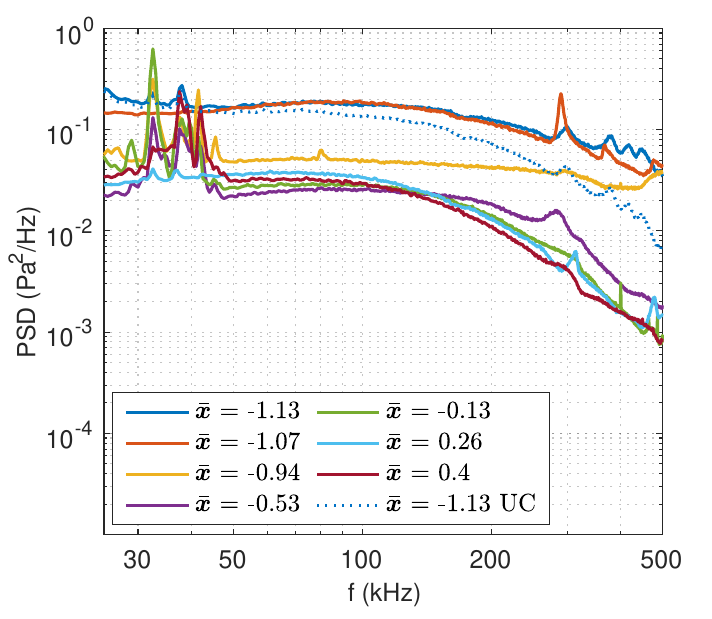}  \label{fig:PSD_PCB_0_deg_Turb}}
		\caption{PSD of pressure measured by PCBs at Mach 5 for cone-slice geometry ($0^{\circ}$ ramp) (a) Laminar case, $Re = 8.6 \times 10^6$ /m, (b) Turbulent case, $Re = 18.8 \times 10^6$ /m. Arrows show evolution of pressure-fluctuations and are discussed in text.}
		\label{fig:PSD_PCB_0_deg_M5}
	\end{figure}

    Figure \ref{fig:PSD_PCB_0_deg_M5} illustrates the power spectral density (PSD) for PCB sensors positioned along the cone-slice geometry at Mach 5 for two distinct $Re$ cases. Consistent sharp peaks below 50 kHz and around 300 kHz, present across all $Re$, are attributed to electronic noise and sensor characteristics \cite{Ort2019}; the latter has been observed in other experiments \cite{Butler2022} using these sensors. The late-laminar case, depicted in figure \ref{fig:PSD_PCB_0_deg_Lam} for $Re = 8.6 \times 10^6$ /m exhibits similarities to the Mach 8 laminar/transitional data detailed in \cite{Pandey2024}. High-frequency content on the cone, indicative of the second-mode instability (approximately 280 kHz, solid arrow), is suppressed along the slice. This suppression is accompanied by the emergence of progressively lower-frequency peaks (dashed arrow), consistent with the amplification of new, lower-frequency second-mode waves within the thickened boundary layer post-expansion \citep{Butler2021ExiF}. In contrast, the turbulent case, shown in Figure \ref{fig:PSD_PCB_0_deg_Turb}, demonstrates boundary-layer fluctuations suppressed from their magnitude prior to the expansion corner. A continuous decrease in the high-frequency regime ($>200$ kHz) is observed up to $\bar{x} = -0.13$, beyond which the spectra become largely similar. The Corcos correction \citep{Corcos1963} was applied to the turbulent spectra to mitigate attenuation effects from the finite size of the PCB sensors. For the cone sensor located at $\bar{x} = -1.13$, the uncorrected PSD, labeled as UC, is also provided in figure \ref{fig:PSD_PCB_0_deg_Turb} for comparison.

\begin{figure} [hbt!]
		\centering
		\subfigure[]
		{\includegraphics[trim={130 240 150 240}, clip, width=0.49\textwidth]{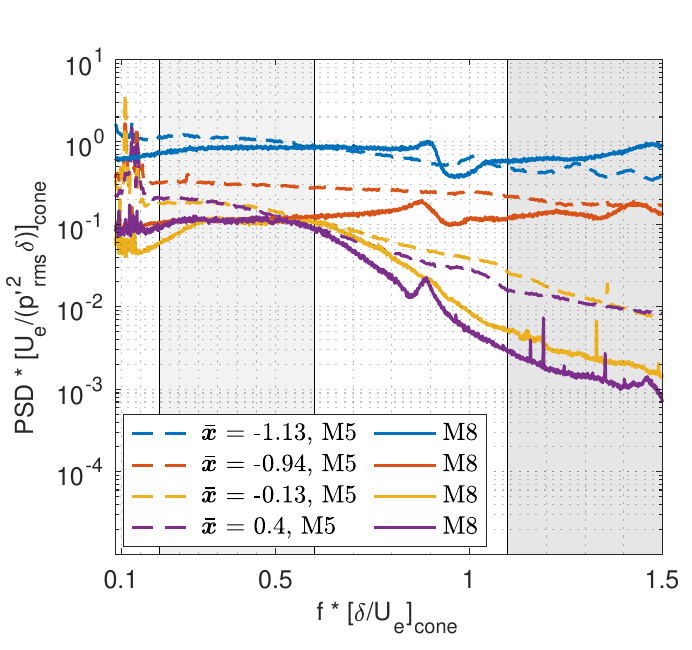} \label{fig:PCB_PSD_0deg_M5vsM8_PSD}}
		\subfigure[]
		{\includegraphics[trim={130 240 150 240}, clip, width=0.49\textwidth]{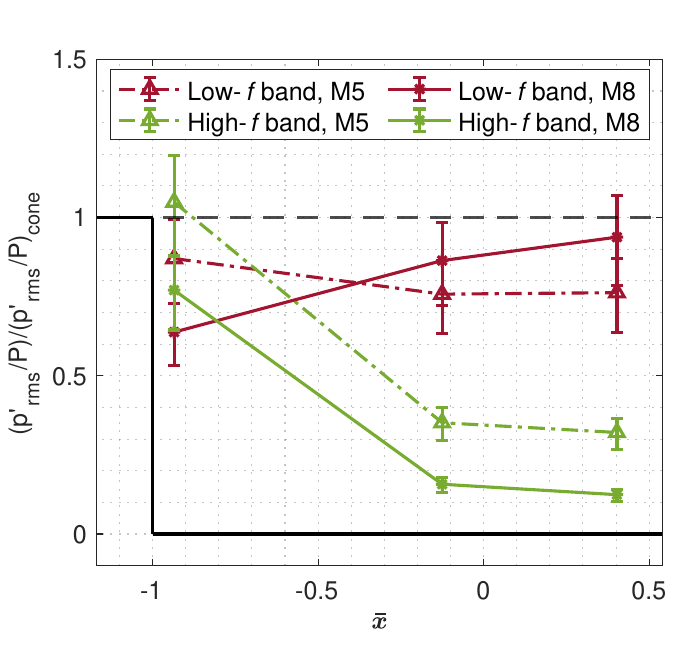}  \label{fig:PCB_rms_0deg_M5vsM8_RMS}}
		\caption{Comparison of Turbulent PCB data at Mach 5 and Mach 8 for the cone-slice geometry ($0^{\circ}$ ramp) (a) Normalized PSD using outer scaling of the boundary layer on the cone, dashed lines are Mach 5 at $Re = 18.8 \times 10^6$ /m and solid lines are Mach 8 at $Re = 16.5 \times 10^6$ /m. (b) Normalized RMS Pressure computed in the two frequency bands bounded in (a) by dotted and dashed black lines.}
		\label{fig:PSD_PCB_0_deg_M5vsM8}
	\end{figure}

	Figure \ref{fig:PCB_PSD_0deg_M5vsM8_PSD} compares the normalized PSD of turbulent boundary layer fluctuations at Mach 5 and 8, with normalization achieved using outer-scaling variables and corrected via the Corcos method. For consistency, measurements upstream of the slice have been used for normalization. Boundary-layer thickness $\delta$ and edge velocity $U_{e}$ were obtained from the mean schlieren images and Taylor-Maccoll equations, respectively. The root-mean-square (RMS) pressure fluctuation ($p\mathrm{'_{rms}}$) for the cone PCB sensor ($\bar{x} = -1.13$) was calculated by integrating fluctuations from 50 to 500 kHz.
    
    The spectra reveal a significant reduction in boundary-layer fluctuations across the expansion, particularly in the high-frequency range, with this attenuation being more pronounced at Mach 8.  To quantify this reduction, RMS pressure was evaluated within the two frequency bands indicated by shaded regions in figure \ref{fig:PCB_PSD_0deg_M5vsM8_PSD} and is presented in figure \ref{fig:PCB_rms_0deg_M5vsM8_RMS}. To account for the decrease in mean pressure along the slice, data were normalized by the local mean pressure from co-located Kulite sensors, yielding a metric of local unsteady fluctuations. Black solid line going from 1 to 0 provide visual reference to the scenario when the fluctuations are completely eliminated across the expansion. The normalized RMS fluctuations in the lower-frequency band show only a modest decrease compared to the upstream boundary layer. Conversely, normalized high-frequency fluctuations dramatically reduce to approximately $30\%$ and $10\%$ of the upstream values at Mach 5 and 8, respectively.

	Prior experimental \citep{Lu1992, Dawson1994, Arnette1995}, computational \citep{Sun2017} and theoretical \citep{Stollery1974} studies on two-dimensional expansion corners demonstrated that the post-expansion recovery length scales with the upstream boundary-layer thickness. Given that the incoming Mach 8 boundary layer is thicker and that it undergoes a more severe expansion, it is not surprising that the recovery of turbulence and the development of secondary flows are limited at Mach 8 as compared to Mach 5. This evolving boundary layer downstream of the expansion determines the characteristics of the shock-boundary-layer interaction (SBLI) on the subsequent ramp geometry.

	\section{Mean flow organization on the cone-slice-ramp geometry}
	\label{subsection:MeanFlowCSR}

	The previous section presented measurements for the expanded boundary layer on the cone-slice geometry. In this section, the focus is on the complete cone-slice-ramp geometry with a $30^{\circ}$ ramp shown in figure \ref{fig:FigGeometry}. Both the Mach number and the $Re$ were varied to understand the effect of the flow parameters on the mean aspects of the resulting SBLI. Table \ref{tab:BLstate} provides approximate $Re$ boundaries for the different SBLI regimes across the two Mach numbers. The flow regimes are classified based on the location of the transition front relative to the SBLI at the slice-ramp junction. In a laminar SBLI, the breakdown to turbulence occurs downstream of the shear-layer reattachment on the ramp. A transitional SBLI occurs when transition is localized within the separation bubble—between the separation and reattachment points—or sufficiently near the separation point such that the incoming boundary layer is already transitional. Within this Reynolds number range, the transition front typically resides on the slice. Finally, a turbulent SBLI is characterized by a transition front located sufficiently upstream of the cone-slice junction; although the boundary layer approaching from the cone is effectively turbulent, the relaminarization effects at the slice junction may suppress this turbulence before it reaches the interaction region. The overall flow field is first qualitatively discussed using schlieren, oil-flow visualization (OFV) and temperature-sensitive paint (TSP) images. Quantitative data from surface heat-transfer and mean-pressure sensors are then presented. These results establish the large-scale structure of the separated flow region and the impingement of the shear layer on the compression ramp. 

\begin{table}[hbt!]
\caption{\label{tab:BLstate} Reynolds number ranges for the different SBLI regimes. $Re$ values are in $\times 10^6$ /m}
\centering
\begin{tabular}{lcccc}
\hline
Mach    &  Laminar  &  	Transitional  &  	Turbulent \\\hline
5       &  $<7.5$   &	$7.5-11.5$ 	  & 	$>11.5$   \\  
				8  		&  $<6$     &	$6-9.5$ 	  & 	$>9.5$ 	  \\
\hline
\end{tabular}
\end{table}

	\subsection{Variation of mean separation}
	\label{subsection:ScaleofSep}

\begin{figure} [hbt!]
		\centering
		\subfigure[]
		{\includegraphics[trim={140 140 140 150}, clip, width=0.49\textwidth]{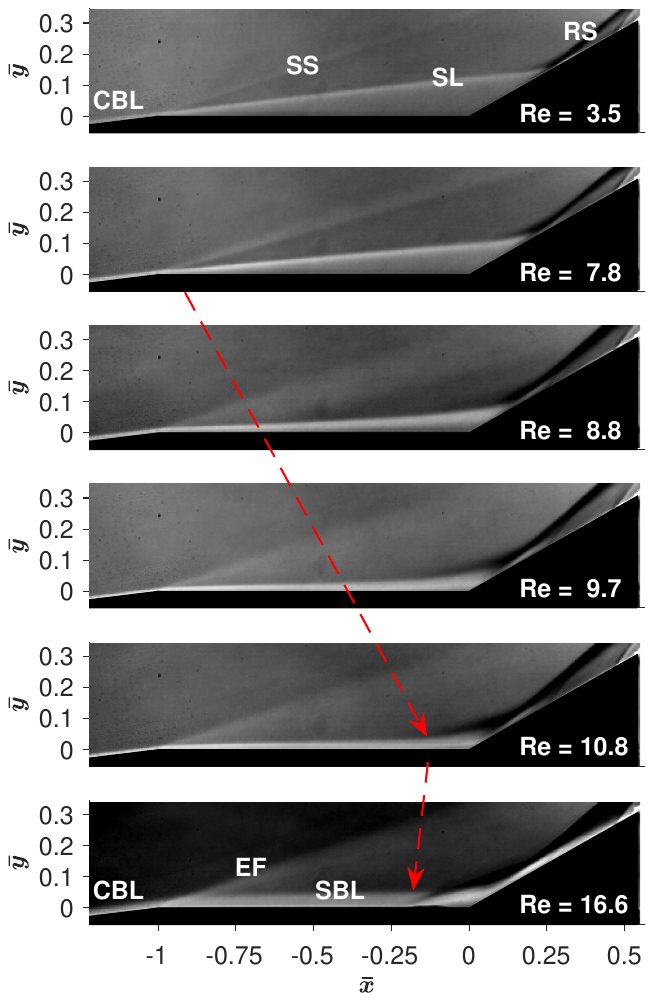} \label{fig:fig_Schlieren_M5}}
		\subfigure[]
		{\includegraphics[trim={140 140 140 150}, clip, width=0.49\textwidth]{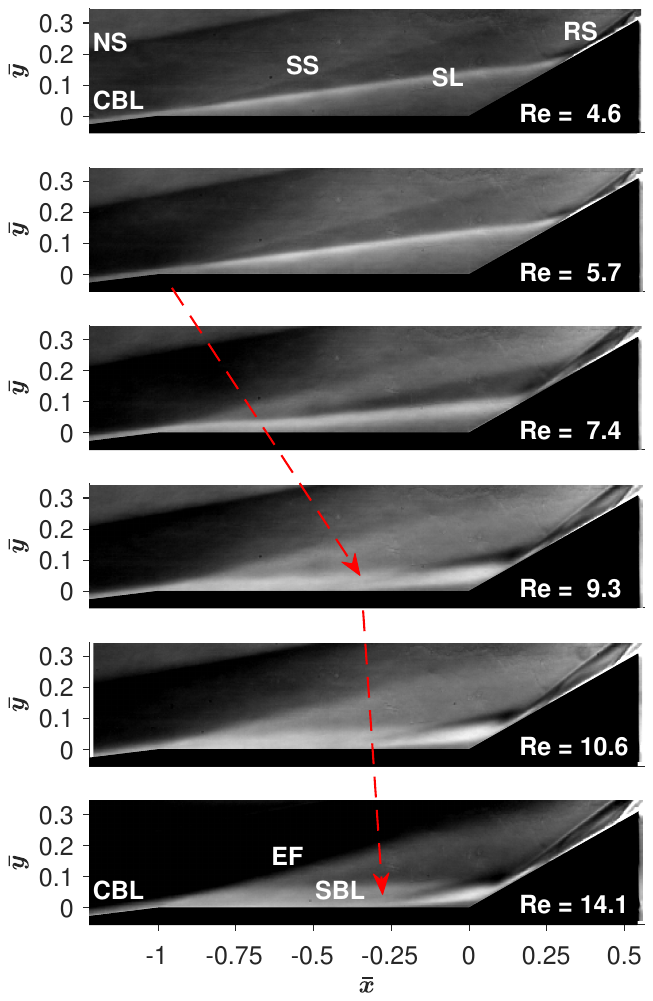} \label{fig:fig_Schlieren_M8}}
		\caption{Temporally-averaged schlieren image at (a) Mach 5 and (b) Mach 8 for the cone-slice-ramp geometry with $30^{\circ}$ ramp. $Re$ values are $= \times 10^6$ /m. Red dashed arrow annotates movement of the mean-separation location with Re. Annotations of SBLI features are described in the text.}
		\label{fig:fig_Schlieren_M5M8}
    \end{figure}

\begin{figure} [hbt!]
		\centering
		\subfigure[]
		{\includegraphics[trim={150 150 170 150}, clip, width=0.45\textwidth]{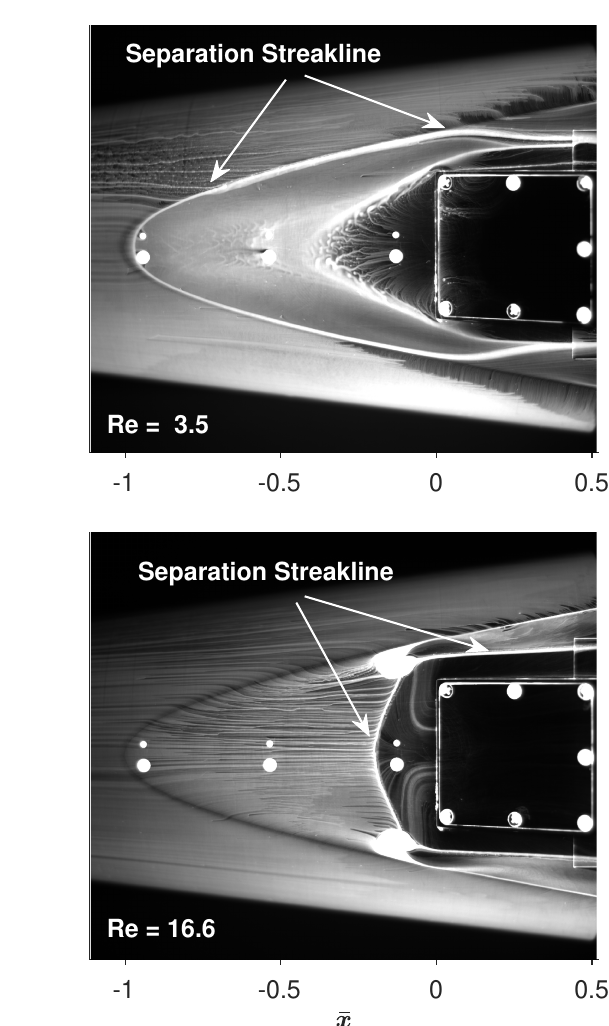} \label{fig:OFV_30deg_M5}}
		\subfigure[]
		{\includegraphics[trim={150 150 170 150}, clip, width=0.45\textwidth]{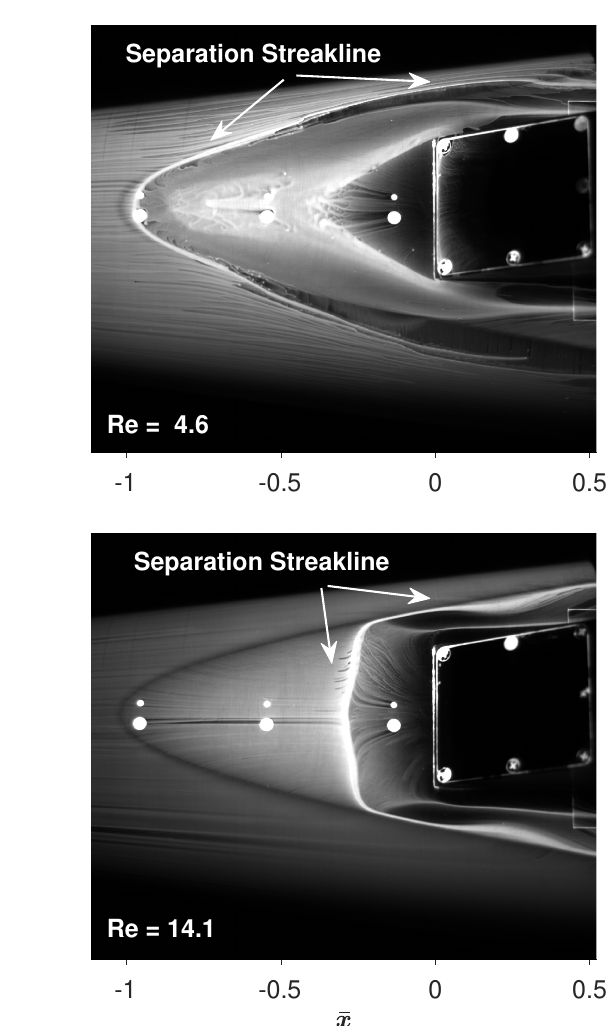} \label{fig:OFV_30deg_M8}}
		\caption{Surface streaklines from oil-flow for the cone-slice-ramp geometry with $30^{\circ}$ ramp at (a) Mach 5 and (b) Mach 8. $Re$ values are $= \times 10^6$ /m.}
		\label{fig:OFV_30deg_M5M8}
    \end{figure}

\begin{figure}
        \centering
        \subfigure[]
        {\includegraphics[trim={160 0 165 0}, clip, width=0.45\textwidth]{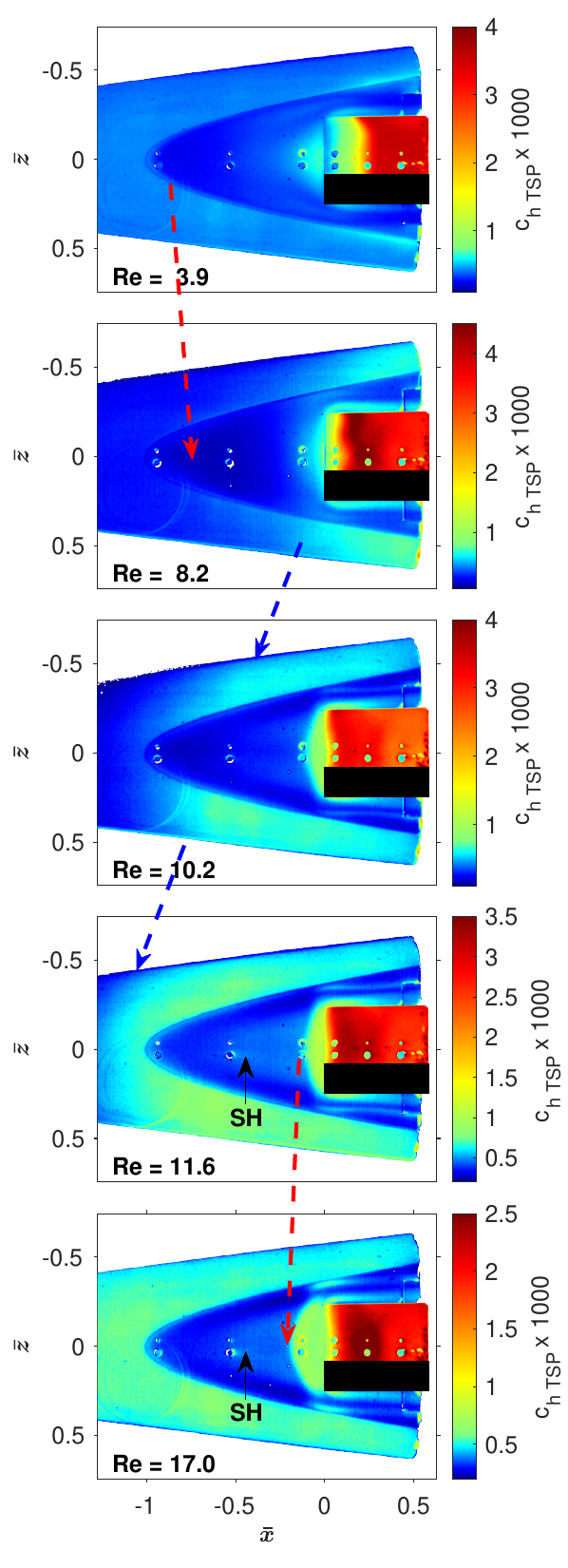} \label{fig:TSP_30deg_M5}}
        \subfigure[]
        {\includegraphics[trim={160 0 165 0}, clip, width=0.45\textwidth]{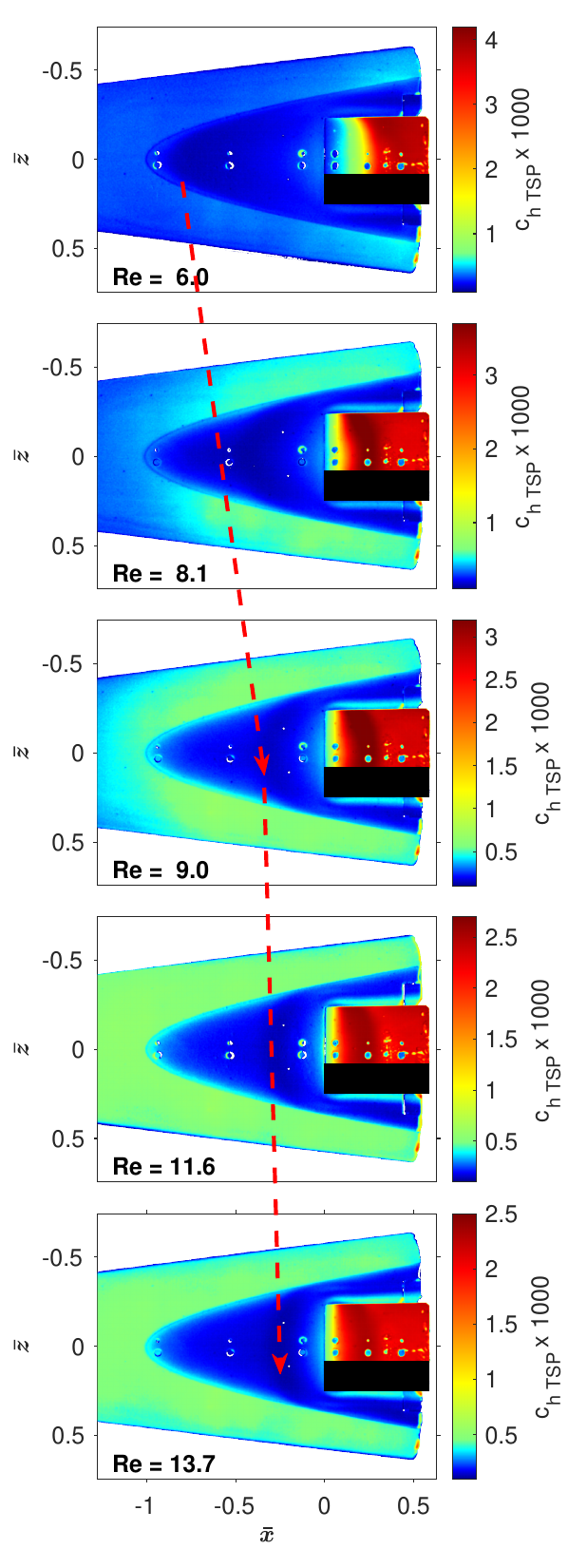} \label{fig:TSP_30deg_M8}}
        \caption{Heat flux derived from TSP for the cone-slice-ramp geometry with $30^{\circ}$ ramp at (a) Mach 5 and (b) Mach 8. $Re$ values are $= \times 10^6$ /m. TSP data was unavailable in the region masked with a black rectangle. Annotations are discussed in the text.}
        \label{fig:TSP_30deg_M5M8}
    \end{figure}

    Figure \ref{fig:fig_Schlieren_M5M8} compares the mean schlieren images (temporally averaged over 0.3 seconds) at Mach 5 and 8. The $Re$ values increase from the top to the bottom and are noted in the individual sub-figures. The corresponding OFV images for the low- and high-$Re$ cases are presented in figure \ref{fig:OFV_30deg_M5M8}; due to differences in the camera orientation across the two Mach number experiments, the field of views are slightly different. Also, Y-axis is not provided because the image plane was not parallel to the slice. Stronger image intensities are produced in regions where the streaklines converge and no lines are visible on the ramp due to both elevated temperatures that can evaporate oil as well as a diverging tendency of the streamlines near reattachment. OFV for the transitional cases have not been shown because the oil streaklines were unable to equilibrate over the course of a HWT run for those cases. Finally, the heat-flux coefficients derived from the TSP images are presented in figure \ref{fig:TSP_30deg_M5M8}; $Re$ values again increase from the top to bottom for each Mach number. The blue hyperbolic region signifies reduced heating due to expansion on the slice, while the yellow-red rectangular region indicates increased heating on the ramp. A black rectangle masks a region on the ramp where TSP data were not acquired. TSP measurements were made in a separate wind-tunnel test campaign with slightly different $Re$ values; therefore, data from the closest $Re$ cases are presented for comparison. Additional TSP data at other $Re$, including line profiles, can be found in \cite{PandeyST2020}.  	

	For the low Reynolds number cases ($Re < 6\times 10^6$ /m), the incoming boundary layer is laminar at both Mach 5 and 8. Schlieren images show that the cone boundary layer (annotated as CBL in figure \ref{fig:fig_Schlieren_M5M8}) separates at the cone-slice corner itself, generating both a separation shock (SS) and a separated shear layer (SL). An expansion fan (EF) also originates at this corner at higher $Re$ when the separation moves downstream. The separated shear layer has a long travel length and reattaches approximately halfway down the ramp where a reattachment shock (RS) forms. The shallower nose shock (NS) is also visible in the field of view of the Mach 8 schlieren. 
    
    Increasing $Re$ within the laminar regime ($Re \approx 3-5\times 10^6$ /m at Mach 5) did not alter the separation location; the expansion corner appeared to `lock-in' the upstream extent of the separation bubble, initiating separation well before the compression ramp is reached. This behavior deviates from typical laminar SBLI without an expansion corner, where experiments have shown slight variation in bubble size with increasing $Re$ \citep{Miller1964, Katzer1989}. The lock-in phenomenon has been studied in two-dimensional supersonic impinging SBLIs near an expansion corner \cite{Chew1979,Chung1995,Guo2024}. While these studies focused on turbulent boundary layers, they demonstrated that shock impingement immediately downstream of an expansion corner can cause the separation front to lock in to the corner. This occurs because the reduced pressure downstream of the expansion enlarges the separation bubble relative to cases without an expansion. Conversely, when the impingement location is further downstream, the separation bubble grows larger but is unable to reach the corner. The present study observes this behavior through variations in ramp angle. For smaller angles ($10^{\circ}$ or $20^{\circ}$), the central core of the separated flow does not extend to the expansion corner, which is consistent with recent CCF findings \cite{Benitez2023, Benitez2025} for flare angles limited to  $12^{\circ}$. However, for larger angles ($30^{\circ}$ or $40^{\circ}$), the lock-in behavior occurs across both Mach 5 and 8 regimes and across the investigated laminar $Re$ range. While this study focuses on the $30^{\circ}$ ramp geometry, the other cases are briefly discussed in section \ref{subsection:UnsteadyLam}.

    For the laminar Mach 5 case, shown in the top subfigure of figure \ref{fig:OFV_30deg_M5}, the line of convergence aligns with the slice contour indicating that the separation is locked-in to the three-dimensional expansion corner. In contrast, the laminar Mach 8 case, shown in the top subfigure of figure \ref{fig:OFV_30deg_M8}, exhibits a separation region that locks into the expansion corner over the central core of the flow but not at the spanwise edges. Further downstream, but still before the compression ramp is reached, the separation bubble extends over the expansion corner onto the cone surface (as outlined by the converging streaklines) implying a larger separation bubble. Closer examination of the schlieren images (compare top subfigures in figures \ref{fig:fig_Schlieren_M5} and \ref{fig:fig_Schlieren_M8}) and TSP images (compare top subfigures in figure \ref{fig:TSP_30deg_M5M8}) confirms that the shear-layer reattachment location for Mach 8 is further downstream on the ramp. The three-dimensional topology of the separation bubble on the present cone-slice-ramp geometry contrasts sharply with the uniform separation observed on axisymmetric bodies at zero angle of attack \cite{Lugrin2022, Davami2025, Benitez2025}.

    As the freestream $Re$ is increased, the cone boundary layer away from the slice-ramp region begins to breakdown to turbulence. The transition front initially manifests at the aft end of the cone and progressively shifts upstream as $Re$ increases. As shown in figure \ref{fig:TSP_30deg_M5M8}, this front appears as a blue-to-green color change on the cone surface (annotated with blue dashed arrows for Mach 5 data), indicating enhanced heating characteristic of a turbulent boundary layer. Within the TSP field of view, the average transition front can be seen moving upstream as $Re$ increases from $8-12 \times 10^6$ /m at Mach 5 and $6-10 \times 10^6$ /m at Mach 8. Detailed analysis of the upstream evolution of transition on a straight $7^{\circ}$ cone in HWT is available in \cite{Casper2016}. 
    
\begin{figure}
        \centering
        \includegraphics[trim={150 260 165 260}, clip, width=0.45\textwidth]{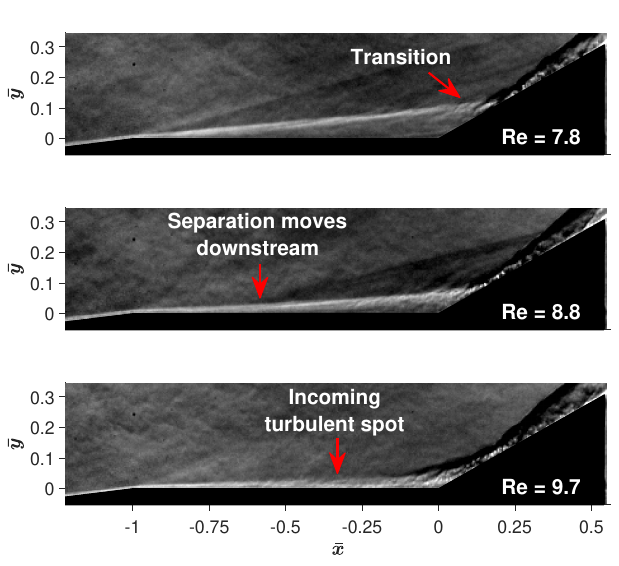} \label{fig:TSP_30deg_M5}
        \caption{Instantaneous schlieren images in the transitional regime at Mach 5. $Re$ values are in $10^6 /$m.}
        \label{fig:SchlierenInst_30deg_M5}
    \end{figure}

    Within the slice-ramp region, transition to turbulence is expedited by additional instability mechanisms present in the separated flow (further detailed in section 5). In the early transitional regime, $Re \approx 8 \times 10^6$ /m at Mach 5, instability waves in the incoming boundary layer have become strong but evidence of turbulent spots is not yet found. Here, transition occurs near reattachment as shown using an instantaneous schlieren image in figure \ref{fig:SchlierenInst_30deg_M5}. A relative decrease in the mean separation size is observed in both mean schlieren and TSP images. The TSP images specifically show a band of increased heating due to transitional reattachment that moves closer to the compression corner. Mach 8 data demonstrates analogous behavior, albeit at an earlier $Re$ of approximately $6.5\times 10^6$ /m.

    Beyond this point, increasing the $Re$ induces a more upstream breakdown of the shear layer; this is evident in the instantaneous schlieren image for $Re = 8.8 \times 10^6$ /m case in figure \ref{fig:SchlierenInst_30deg_M5}. This breakdown collapses the separation region over a short $Re$ range, leading to a dramatic reduction in separation bubble size at both Mach numbers. Figure \ref{fig:fig_Schlieren_M5M8} uses a dashed-red arrow to indicate the downstream shift of the separation location, with corresponding upstream movement of reattachment locations. At this point, a post-expansion boundary layer is found on the slice (SBL). This rapid reduction persists until the transition front advances upstream of the mean separation location. For Mach 5, this occurs at $Re \approx 10-11\times 10^6$ /m, as evidenced by the shear stress and TSP data. At Mach 8, this occurs at a lower $Re$ of approximately $9\times 10^6$ /m. At this stage, the Mach 5 separation region reaches its minimum size, which is considerably smaller than that observed at Mach 8. 

    As $Re$ increases, the transition front of the cone boundary layer shifts further upstream, causing it to exhibit increasingly turbulent characteristics as it enters the slice region. In this regime, the SBLI separation size exhibits an opposite dependence on freestream $Re$ across the two Mach numbers. For Mach 5, the transitional regime concludes with the minimum mean separation size. Subsequent increases in $Re$ result in an enlargement of the separation region, with both separation and reattachment fronts relocating away from the slice-ramp corner. Conversely, for Mach 8, the rapid reduction in separation size observed during the transitional regime becomes a more gradual reduction at higher $Re$ as the boundary layer becomes more turbulent. A dashed red arrow in schlieren and TSP images annotates these trends in the mean separation location.

    The disparate behavior observed in the turbulent regime can be elucidated by analyzing heating patterns on the slice upstream of the ramp. Canonical compression ramp studies \citep{Simeonides1995, Arnal2004} have established that laminar shock-induced boundary-layer separation correlates with a localized decrease in heating, whereas turbulent separations are associated with a localized increase. Transitional cases exhibit an intermediate behavior dependent on boundary-layer intermittency. In the current geometry, the presence of the slice upstream of the ramp introduces additional complexity. The expansion at the slice itself reduces boundary-layer heating, manifesting as the blue hyperbolic region in the TSP images across all $Re$ and both Mach numbers. In the laminar regime, this reduction compounds the heating decrease resulting from laminar separation at the cone-slice corner. As $Re$ increases and the time-averaged separation of the early-transitional SBLI shifts downstream, the localized heating dip also migrates downstream. This phenomenon is annotated in the TSP images with a dashed-red arrow for $Re < 10\times 10^6$ /m.

    At Mach 8, even with the cone boundary layer transitioning to turbulence for $Re > 10\times 10^6$ /m, the separation characteristics surprisingly retain their laminar behavior. Consistent with schlieren observations, the localized region of decreased heating, rather than the expected turbulent increase in heating, shifts downstream with increasing $Re$. This unexpected phenomenon stems from the significant relaminarization effects induced by the expansion corner, which effectively transforms the turbulent interaction into a transitional SBLI, as detailed by \cite{Pandey2024}. 
    
    In contrast at Mach 5, the high $Re$ cases present two distinct heating features on the slice. First, as the transition front advances upstream of the slice, a region of secondary heating appears, offset from the hyperbolic slice (annotated as SH in the TSP images). This is likely due to the boundary layer re-transitioning as it recovers after the expansion. Second, a sharp increase in heating is observed upstream of the ramp. Its location correlates well with the boundary-layer separation seen in schlieren images. Unlike Mach 8, this separation location moves upstream with increasing $Re$, as indicated by a red dashed arrow in both schlieren and TSP images. 

    Holden \cite{Holden1977} used the boundary-layer database compiled by Johnson and Bushnell \cite{Johnson1970} to provide an explanation for the increase in turbulent SBLI size with increasing $Re$. This explanation centers on the velocity exponent of the boundary layer (which can be considered a metric for its fullness and resistance to separation) which exhibits an overshoot at the conclusion of transition. The database indicated that the velocity exponent initially increases, then decreases with $Re$ at the end of the transition process, before gradually increasing in the fully turbulent regime. Holden \cite{Holden1977} posited that turbulent SBLI studies reporting an increase in separation size with $Re$ were conducted in the late-transitional regime where the velocity exponent was decreasing with $Re$. 
    
    The Mach 5 turbulent SBLI results on the current cone-slice-ramp geometry suggest such late-stage transitional behavior rather than equilibrium turbulent behavior. Conversely, Mach 8 cases exhibit the typical transitional trend of decreasing separation size for increasing $Re$. Thus, the expansion corner acts to push the SBLI characteristics to an earlier stage of transition and this effect becomes stronger with Mach number. 

\subsection{Surface sensor measurements: heat transfer and pressure}
	\label{subsection:QandPmean}

\begin{figure}  [hbt!]
		\centering
		{\includegraphics[trim={145 190 160 190}, clip, width=0.52\textwidth]{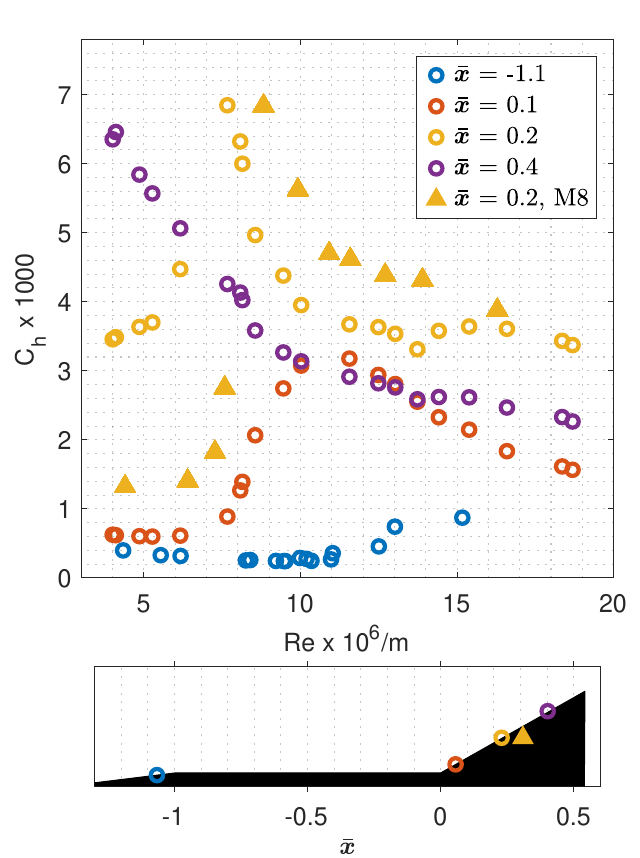}}
		\caption{Heat-flux coefficient at Mach 5 as measured using Schmidt-Boelter gauge for the cone-slice-ramp geometry with $30^{\circ}$ ramp. The bottom subfigure shows the location of the sensors.}
		\label{fig:SB_allRe_M5}
	\end{figure}

    This subsection supplements the qualitative flow field description with quantitative surface sensor measurements acquired near the plane of symmetry. Figure \ref{fig:SB_allRe_M5} presents the Stanton number $(c_h)$ measured by the  Schmidt-Boelter gauges, averaged over the steady portion of the HWT run, spanning the full range of $Re$. Measurements were acquired on the cone upstream of the slice and three ramp locations, where reattachment heating occurs. Mach 8 data, acquired only at the second ramp sensor location, is included in the same plot for comparison. The Stanton number was calculated as $c_h = \dot{q}/[\rho_e U_e C_p (T_{0_e}-T_w)]$. Here, the subscript `$e$' denotes edge quantities derived from Taylor-Maccoll equations, while heat flux $(\dot{q})$, and local wall temperature $(T_w)$ were directly measured by the sensor. 
    
    The cone sensor at $\bar{x} = -1.1$ exhibits very small $c_h$ values which begin to increase within the transitional regime, consistent with figure \ref{fig:SS_0deg_allRe_M5}. Conversely, the sensors on the ramp exhibit non-monotonic behavior with peak heating rates achieved at different $Re$. To first order, these trends are dictated by the relative position of the transition and reattachment fronts with respect to the sensors. As established in two-dimensional SBLI studies \citep{Simeonides1995, Arnal2004}, heating on a compression ramp typically demonstrates an initial sharp increase, peaks downstream of reattachment, and then gradually stabilizes to the local boundary layer values. Similar trends can be observed here as the reattachment front moves on the ramp with varying $Re$.

   At Mach 5, for $Re$ ranging from approximately $4-5.5 \times 10^6$ /m, the SBLI remains laminar with minor variations in the separation size. Reattachment occurs near the sensor at $\bar{x} = 0.2$. Neither this sensor, nor the upstream sensor at $\bar{x} = 0.1$, exhibit significant changes in this $Re$ regime, as they are located beneath the reattaching shear layer. However, the downstream sensor at $\bar{x} = 0.4$ shows a rapid decrease in heating, indicating an upstream shift of the transition and reattachment fronts as $Re$ increases. 

   For $Re > 6 \times 10^6$ /m, the reattaching shear layer begins to break down, prompting an upstream shift in the reattachment. Between $Re$ of $6-8 \times 10^6$ /m, this phenomenon results in a sharp rise in $c_h$ values at $\bar{x} = 0.2$, concurrent with the reattachment heating sweeping over the sensor. The peak $c_h$ at this location is achieved at approximately $Re \sim 7.5-8 \times 10^6$ /m, at which point the reattachment is upstream of this sensor, as evidenced by the corresponding schlieren and TSP images in subsection \ref{subsection:ScaleofSep}. Throughout this $Re$ regime, the downstream sensor at $\bar{x} = 0.4$ continues to show a decreasing trend, as the reattachment front moves progressively upstream. On the other hand, the sensor at $\bar{x} = 0.1$ maintains a constant reading, remaining beneath the separated shear layer. 

   For $Re$ ranging from approximately $8-11 \times 10^6$ /m, the reattachment front progressively shifts upstream between the sensors located at $\bar{x} = 0.1$ and $\bar{x} = 0.2$. Consequently, the aforementioned heating trend is repeated with decreasing $c_h$ at $\bar{x} = 0.2$ and increasing $c_h$ at $\bar{x} = 0.1$. The latter sensor registers peak heating rates when the separation size is at a minimum, occurring at approximately $Re \sim 11 \times 10^6$ /m. Beyond this $Re$, the gradual increase of the separation region mitigates the rapid decline in $c_h$ at $\bar{x} = 0.2$ and $\bar{x} = 0.4$. Concurrently, heating rates at $\bar{x} = 0.1$ decrease as reattachment moves downstream from the sensor. 
   
   Mach 8 data, which was only available at $\bar{x} = 0.2$, exhibits trends analogous to Mach 5. As previously discussed, the larger separation size at Mach 8 in the low-$Re$, laminar regime results in a more downstream reattachment. As a result, the sensor at $\bar{x} = 0.2$ records lower heating rates compared to Mach 5. Notably, with increasing $Re$, a comparable peak heating rate is achieved, although at a slightly higher $Re$ than for Mach 5. The peak is followed by a decrease in values as the reattachment location progressively shifts upstream with increasing $Re$. Despite the three-dimensionality of the present geometry and the relaminarization effects of the expansion corner, the qualitative trends of reattachment heating near the plane of symmetry are similar to canonical two-dimensional flows.
   
    \begin{figure} [hbt!]
		\centering
		\subfigure[]
		{\includegraphics[trim={120 190 140 200}, clip, width=0.49\textwidth]{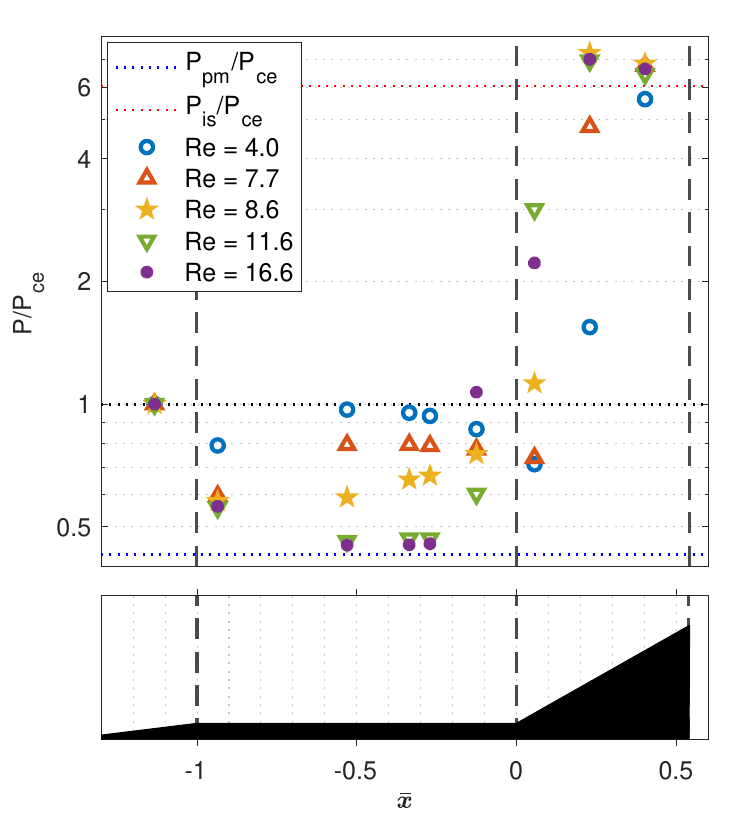} \label{fig:MeanP_M5}}
		\subfigure[]
		{\includegraphics[trim={120 190 140 200}, clip, width=0.49\textwidth]{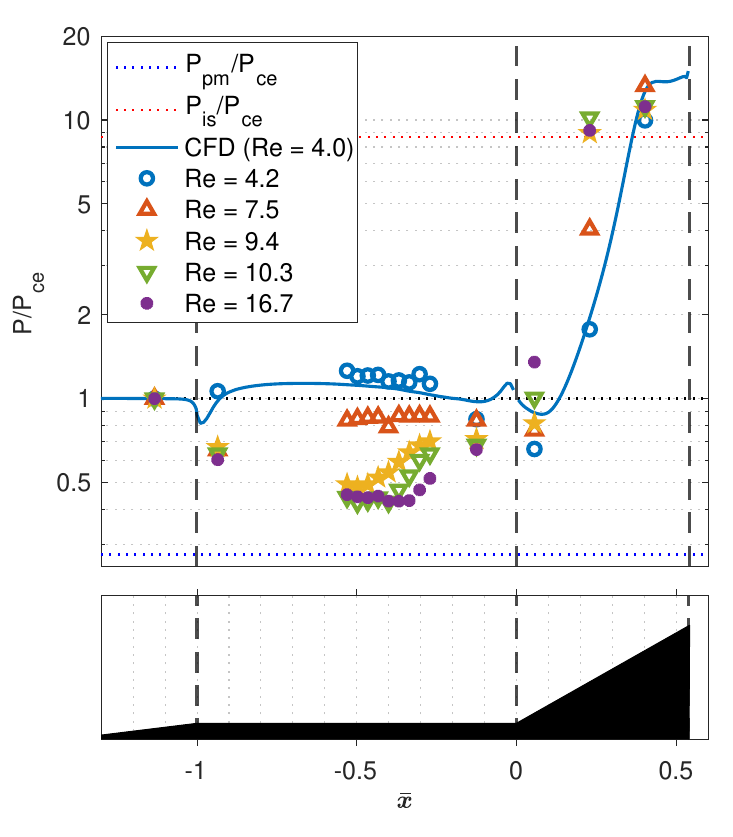} \label{fig:MeanP_M8}}
		\caption{Mean surface pressure along the plane of symmetry compared with inviscid estimates for the cone-slice-ramp geometry with $30^{\circ}$ ramp (a) Mach 5 (b) Mach 8, $Re$ values are $\times 10^6$ /m. CFD data at Mach 8 obtained from \cite{Sadagopan2023}}
		\label{fig:MeanP_M5M8}
	\end{figure}

    The mean pressure measurements at Mach 5 and 8 are presented in figure \ref{fig:MeanP_M5M8} for selected $Re$ cases. Data has been non-dimensionalized by the measured cone-edge pressure $(P_{\mathrm{ce}})$ which closely matches the Taylor-Maccoll equation estimates. A logarithmic $y$-axis accentuates pressure changes across the slice and ramp. Three reference values are included: $P_{\mathrm{ce}}$ (black dotted line) which represents cone-edge pressure measured by the upstream cone sensor, $P_{\mathrm{pm}}$ (blue dotted line) which represents the expected slice pressure from a $7^{\circ}$ two-dimensional Prandtl-Meyer expansion, and $P_{\mathrm{is}}$ (red dotted line) which represents the inviscid ramp pressure derived from two-dimensional oblique-shock relations, using $P_{\mathrm{pm}}$ as the upstream pressure. 
    
    Variations in pressure distributions across the $Re$ range underscore the significant role of viscous effects in SBLIs, which are not captured by inviscid models. Two-dimensional studies have established that pressure distributions exhibit an initial increase near separation, followed by a plateau, before a sharp rise near the corner. Furthermore, the pressure on the ramp may exhibit an overshoot before equilibrating to the inviscid oblique-shock values. In the current geometry, this pressure distribution is superimposed on the pressure decrease induced by the expansion corner.    

    The Mach 8 dataset, where measurement with finer spatial resolution was available, is discussed first. In the laminar regime, flow separation occurs near the cone-slice corner, with pressure sensors on the slice measuring pressures within the separated region. As shown by blue markers ($Re \approx 4 \times 10^6$ /m), the pressure on the slice is nearly equivalent to the cone edge pressure. This is attributed to the compensating effects of expansion at the corner and the compression from the separation shock. Results from computational simulations at Mach 8 from \cite{Sadagopan2023} are also provided for comparison. Both experimental and computational data show a localized pressure dip near the ramp corner at $\bar{x} = 0.1$, a feature indicative of secondary separation in two-dimensional SBLIs \cite{Hao2021}. Downstream along the ramp, pressure increases sharply, approaching the inviscid estimate by the final measurement location. 

    Increasing $Re$ leads to a shrinking separation region and a downstream shift of the separation front as the flow becomes transitional. The non-dimensional mean pressure of the slice sensors, for example at $\bar{x} = -0.94, -0.5$, becomes independent of $Re$ variations once the separation front moves downstream. The plateau pressure downstream of separation depends on the continuously decreasing slice pressure. As a result, as $Re$ increases and separation moves downstream, the plateau pressure decreases. For $Re > 9 \times 10^6$ /m, the mean separation occurs near the middle of the slice where the dense sensor array captures the initial increase in pressure. With the available spatial resolution near the corner, the pressure appears to increase continuously for these cases and it is not clear if a pressure dip similar to the laminar scenario exists at higher $Re$. Along the ramp, the upstream movement of the reattachment location correlates to a faster pressure rise to the inviscid value. 
    
    Mach 5 data exhibit comparable trends along the slice, despite limited sensor resolution. The pressure dip at the slice-ramp corner aligns with observations at Mach 8. Additionally, the plateau pressure decreases with increasing $Re$. Notably, the slice pressure at Mach 5 is significantly lower than at Mach 8, more closely aligning with $P_{\mathrm{pm}}$.   	

    This section provided a detailed analysis on the cone-slice-ramp geometry with a $30^{\circ}$ ramp across two Mach numbers and varying Reynolds number. For laminar Reynolds numbers, the separation locks into the start of the expansion along the centerline. However, there are differences in the three-dimensional flow field between the two geometries in the spanwise direction. Relaminarization has been found to be stronger at Mach 8 than Mach 5, though it is clearly evident in both cases, and this influences the size of the separation region. The reattachment heating line appears to mimic behavior found in the literature for two-dimensional compression ramps. Three-dimensional characteristics are observed at the spanwise edges that deviate from recognized two-dimensional behavior.

\section{Flow unsteadiness on the cone-slice-ramp geometry}
	\label{section:Unsteadiness}

    In this section, several sources of unsteadiness within the SBLI region will be discussed. This unsteadiness has been identified using unsteady surface pressure measurements and high-framerate schlieren data. Since the Mach 8 dataset was acquired with a finer spatial resolution of sensors on the slice, the discussion is primarily focused there. The discussion has been split into four subsections, one each for the laminar, transitional, and the turbulent regimes, and a final section focusing on variations due to Mach number. 

    \subsection{Unsteadiness in the laminar regime}
	\label{subsection:UnsteadyLam}

\begin{figure} 
		\centering
		{\includegraphics[trim={108 180 110 179}, clip, width=0.5\textwidth]{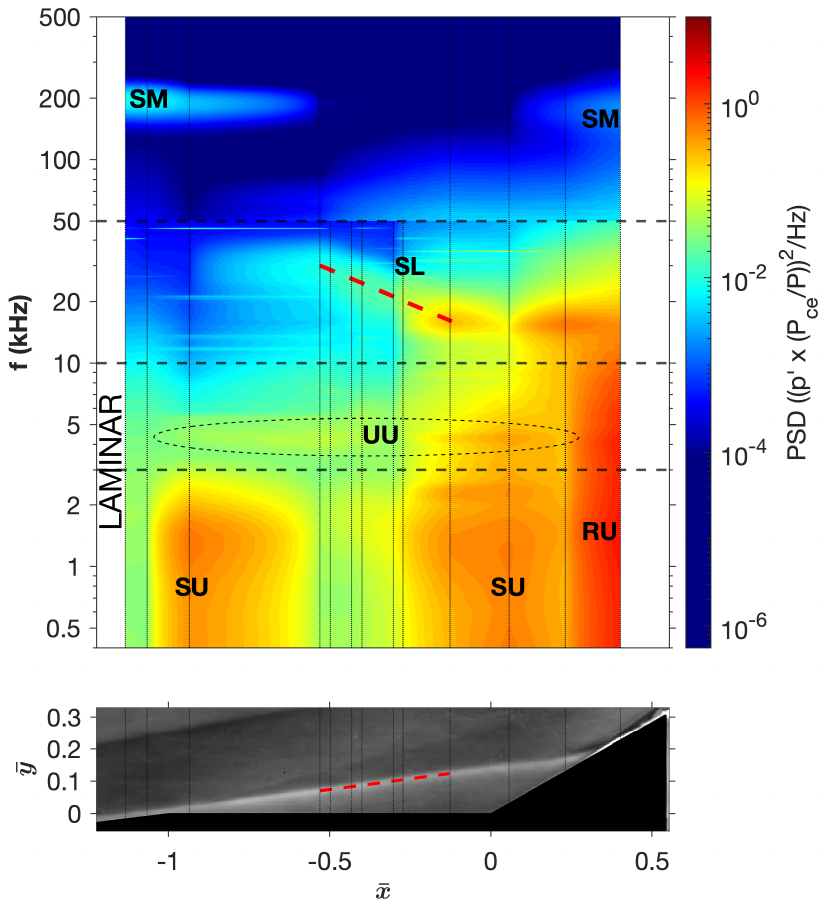}}
		\caption{PSD contours from combined Kulite ($0.4-50$ kHz) and PCB ($50-500$ kHz) data for laminar conditions, Mach 8, $Re \approx 4\times 10^6$ /m, with a corresponding schlieren image (bottom). Dotted vertical lines show the sensor locations and horizontal dashed black lines demarcate the frequency ranges discussed in the text. A dashed red line annotates the shear-layer height and frequencies.}
		\label{fig:PSDcontour_Lam_M8}
	\end{figure}

    The nominally laminar regime nonetheless exhibits several modes of unsteadiness due to presence of instabilities. These fluctuations are registered in the surface pressure sensors and figure \ref{fig:PSDcontour_Lam_M8} shows the combined PSDs of the Kulite and the PCB sensors in a contour form. A corresponding time-averaged schlieren image is shown below the PSD contour for orientation. Dotted-vertical lines mark the locations of the sensors where the data is available, and the PSD contour has been made by interpolating the available data in the spatial dimension; caution must be exercised in interpreting data where the sensor spatial resolution is sparse, such as in the region $\bar{x} = -0.94$ to $\bar{x} = -0.53$. Data up to 50 kHz has been obtained from the Kulite sensors and the data in the 50-500 kHz range has been obtained from the PCB sensors. The PSDs have been scaled by the mean pressure ratio ($P_{ce}/P$ where $P$ is the local mean pressure and $P_{ce}$ is the mean pressure on the cone); this scaling prevents the large fluctuations near reattachment from dominating the PSD contour. Un-normalized line plots of PSDs can be obtained from \cite{PandeyST2022}.
    
    The fluctuations measured by the surface pressure sensors near the plane of symmetry demonstrate a number of temporally and spatially localized regions of unsteadiness; horizontal black-dashed lines have been used to demarcate the different frequency bands where each of these different modes of unsteadiness are active. The high-frequency content ($> 100$ kHz) is primarily due to the existence of second-mode  instability waves (annotated as SM in figure \ref{fig:PSDcontour_Lam_M8}) at about 200 kHz that are present in the incoming laminar boundary layer and then get carried along the separated shear layer. As distance between the separated shear layer and the surface sensors increases, the fluctuations induced by the propagating second-mode waves are not perceived by the sensors, however, off-body focused-laser differential interferometry (FLDI) measurements have confirmed their existence in the separated shear layer \citep{PandeyST2022, Benitez2023}. The second-mode signature is picked up by the downstream sensors on the ramp where the shear layer reattaches and a slight decrease in peak frequency from 200 kHz to 180 kHz is observed. 
	
    \begin{figure} [hbt!]
		\centering
		{\includegraphics[trim={0 145 0 155}, clip, width=0.95\textwidth]{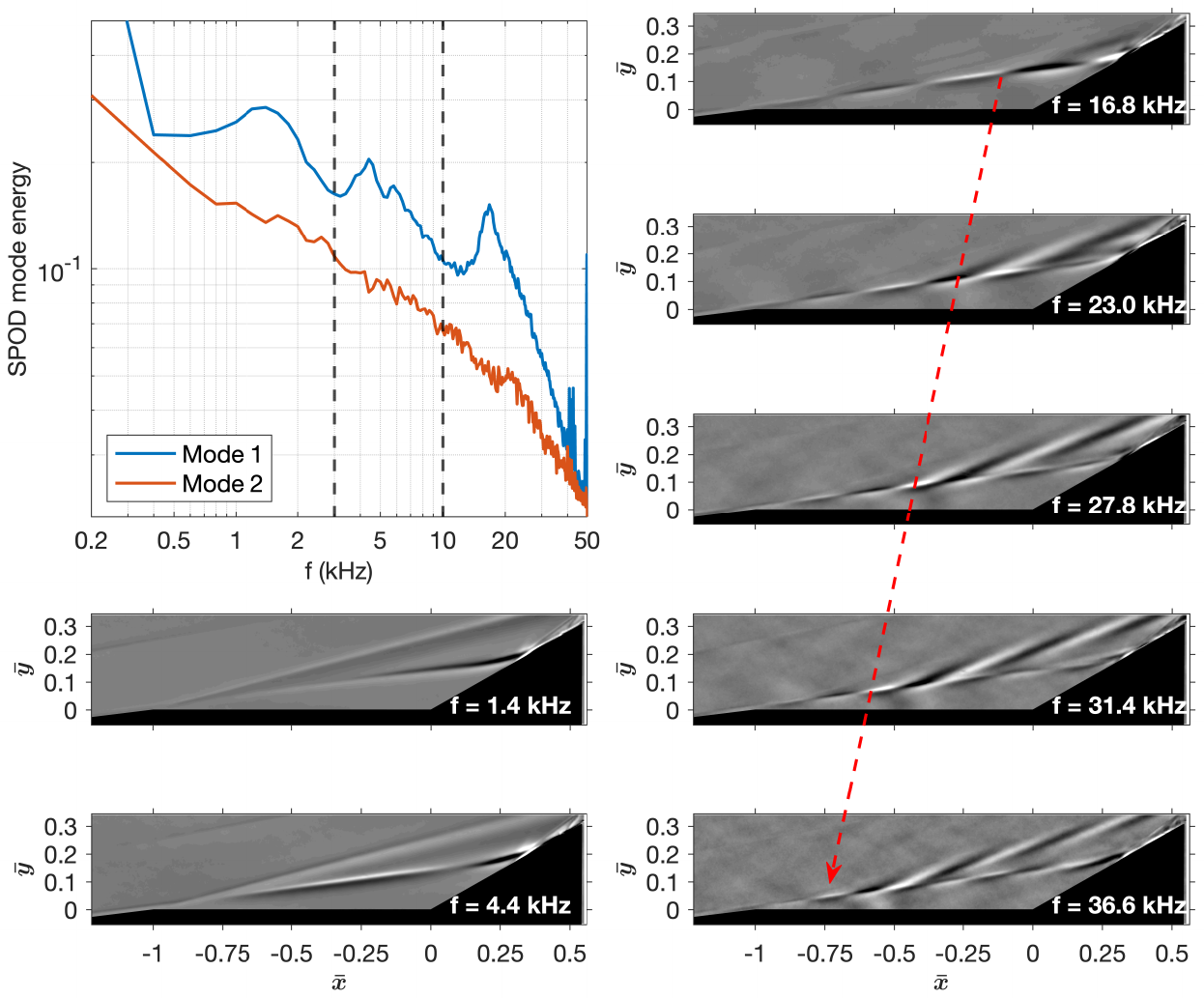}}
		\caption{Schlieren SPOD for Mach 8, $Re = 4.3 \times 10^6$ /m. Dashed lines denote demarcations between frequency bands of analysis.}
		\label{fig:Schlieren_SPOD_Lam_M8}
	\end{figure}

    In the low- to mid- frequency range, three frequency bands have been identified -- 0-3 kHz, 3-10 kHz, and 10-50 kHz. To understand the motion that corresponds to these different modes of unsteadiness, spectral analysis of the schlieren data was performed. Figure \ref{fig:Schlieren_SPOD_Lam_M8} shows the real part of the spectral proper orthogonal decomposition (SPOD) modes for 1 second of schlieren images acquired at 100 kHz. Ample separation between the first two SPOD modes at most frequencies suggests that meaningful structures have been computed as the eigenfunctions \citep{Towne2018, Schmidt2020}. In previous work \cite{PandeyST2022}, schlieren measurements at 200 kHz and 875 kHz were acquired and no differences in the modal distribution were observed. Furthermore, SPOD energy is found to be distributed in the same frequency bands as the pressure bands mentioned above. As such, SPOD-derived eigenmodes reveal the dominant spatiotemporal motions that contribute to the fluctuating energy in the surface pressure and schlieren measurements.  

    Starting with the lowest frequency band, figure \ref{fig:PSDcontour_Lam_M8} shows that unsteady pressure fluctuations exist in the 0-3 kHz band (peak at around 1.4 kHz), at the separation point, slice-ramp corner and the reattachment locations. The SPOD eigenmode at 1.4 kHz is shown in figure \ref{fig:Schlieren_SPOD_Lam_M8} and demonstrates a large-scale, bubble-breathing motion where the motions of the shear layer and the separation shock are strongly coupled. These are marked in figure \ref{fig:PSDcontour_Lam_M8} as SU for separation unsteadiness. The two move in tandem such that as the size of the bubble increases, it outwardly deforms the shock, and vice-versa. Unlike canonical compression ramps in which the shock foot translates with the separation bubble breathing (e.g., \cite{Clemens2014}), here the shock foot and separation point remain locked to the expansion corner, but the shape of both deform as the separation bubble breathes. The pressure sensor located at $\bar{x} = -0.94$ is close to the separation-shock foot and thus picks up this low-frequency shock motion. This unsteadiness also registers near the slice-ramp corner, which suggests that the separation bubble couples into this large-scale motion. The sensor at $\bar{x} = 0.4$ lies downstream of reattachment and indicates that the multiple low-frequency bands have merged into a single range of elevated fluctuations, suggesting that at this point broadband turbulence has emerged. This portion is marked as RU in figure \ref{fig:PSDcontour_Lam_M8} to highlight the reattachment unsteadiness.
    
    In the next frequency band, a relatively weak pressure fluctuation near 4.4 kHz is observed in the upstream part of the slice ($\bar{x} = -0.94, -0.5$) and a narrow-band peak is seen near the slice-ramp corner ($\bar{x} = 0.1$). These features are annotated with a UU to highlight the presence of this unknown unsteadiness. The 4.4 kHz peak is confirmed in the SPOD modal energy plot. The corresponding eigenmode demonstrates braiding of the shear layer that suggests convecting shear-layer undulations and coupled separation-shock movement. The wavelength of the undulation is approximately the same as the streamwise extent of the shear layer. The physical mechanism responsible for the observed 4.4 kHz spectral peak is unknown and merits further study. 

    Finally, the 10-50 kHz band is associated with shear-layer unsteadiness (SL) that evolves along the length of the bubble. Figure \ref{fig:PSDcontour_Lam_M8} shows a range of unsteady frequencies annotated with a red dashed line and annotated as SL; these are first observed at about 32 kHz at $\bar{x} = -0.5$, which is the first sensor located within the separated shear layer, and decrease monotonically to about 17 kHz by $\bar{x} = -0.1$. As the PSD contour shows, the strength of the unsteadiness increases over this streamwise distance as does the height of the shear layer which has also been annotated with a red, dashed line in the time-averaged schlieren image. Beyond $\bar{x} = -0.1$, the interaction with the ramp arrests the decrease in the peak shear-layer frequency, and the shear-layer reattaches on the ramp with a flapping frequency of about 17 kHz. The peak amplitude increases at $\bar{x} = 0.2$ which is close to reattachment. 

    The schlieren SPOD modal energy in figure \ref{fig:Schlieren_SPOD_Lam_M8} peaks at about 17 kHz corresponding to the frequency of the reattachment unsteadiness of the shear layer. The eigenmodes corresponding to several frequencies in the 17-37 kHz range are included in figure \ref{fig:Schlieren_SPOD_Lam_M8} and demonstrate the spatio-temporal character of the shear-layer unsteadiness. The eigenmodes show shear-layer braiding which is representative of shear-layer undulations induced by convecting structures. The wavenumber of the undulations increases with frequency suggesting a fairly constant wave speed of convection. In addition, the streamwise location of the maximum undulation moves upstream with increasing frequency (annotated using a dashed red arrow) and corresponds well with the location and the peak frequency measured by the corresponding surface-pressure sensors. Furthermore, strong radiation of energy is observed as the shear layer deforms into the hypersonic freestream with the radiation initiating at the maximum amplitude location. 

    \begin{figure} 
		\centering
		\subfigure[]
		{\includegraphics[trim={150 300 150 300}, clip, width=0.49\textwidth]{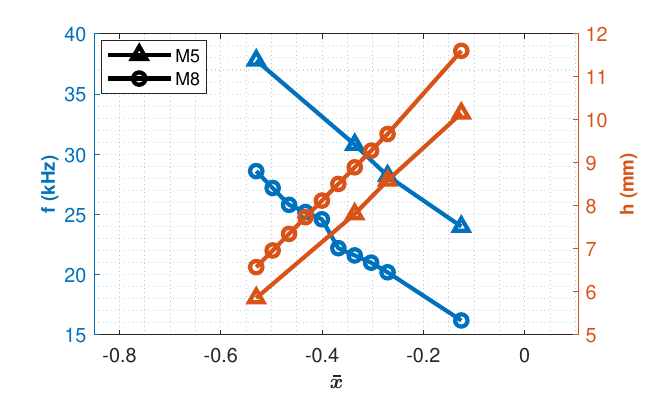} \label{fig:shearlayer_fvsh_M5M8}}
		\subfigure[]
		{\includegraphics[trim={150 300 150 300}, clip, width=0.49\textwidth]{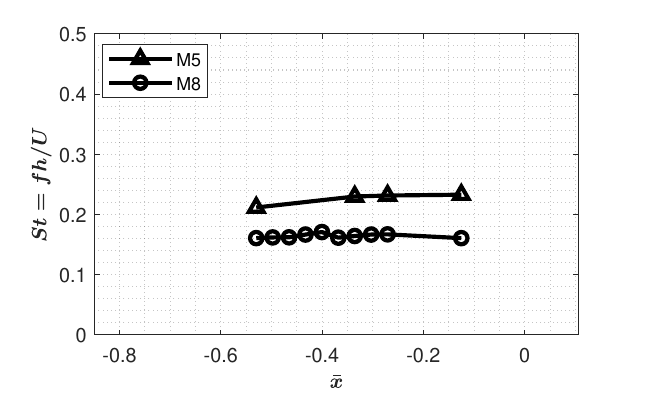} \label{fig:shearlayer_St_M5M8}}
		\caption{(a) Evolution of mid-band peak frequency (f) obtained from Kulite PSDs and shear-layer height (h) obtained from averaged schlieren data. (b) Strouhal number evaluated using f, h, and edge-velocity. $Re \approx 4\times 10^6$ /m for both Mach 5 and Mach 8 data.}
		\label{fig:ShearLayer_fvsh_St_M5M8}
	\end{figure}
    
   Upstream of the ramp, the peak frequency of the shear-layer unsteadiness scales inversely with the shear-layer height. This phenomena is demonstrated in figure \ref{fig:shearlayer_fvsh_M5M8} where the peak frequency has been obtained from the surface-pressure sensors and the shear-layer height has been obtained from the time-averaged schlieren image. Corresponding Mach 5 data is also shown and like the Mach 8 dataset, it shows monotonically-decreasing peak frequencies in the streamwise direction. Figure \ref{fig:shearlayer_St_M5M8} shows that a Strouhal number based on shear-layer height scales the peak frequencies. Here, cone-edge velocity determined from Taylor-Maccoll equations for a straight cone has been used as the velocity scale. These values are approximately 1050 m/s and 1220 m/s, for the Mach 5 and 8 laminar cases, respectively. 
	
	\begin{figure} [hbt!]
		\centering
		{\includegraphics[trim={25 220 40 220}, clip, width=0.98\textwidth]{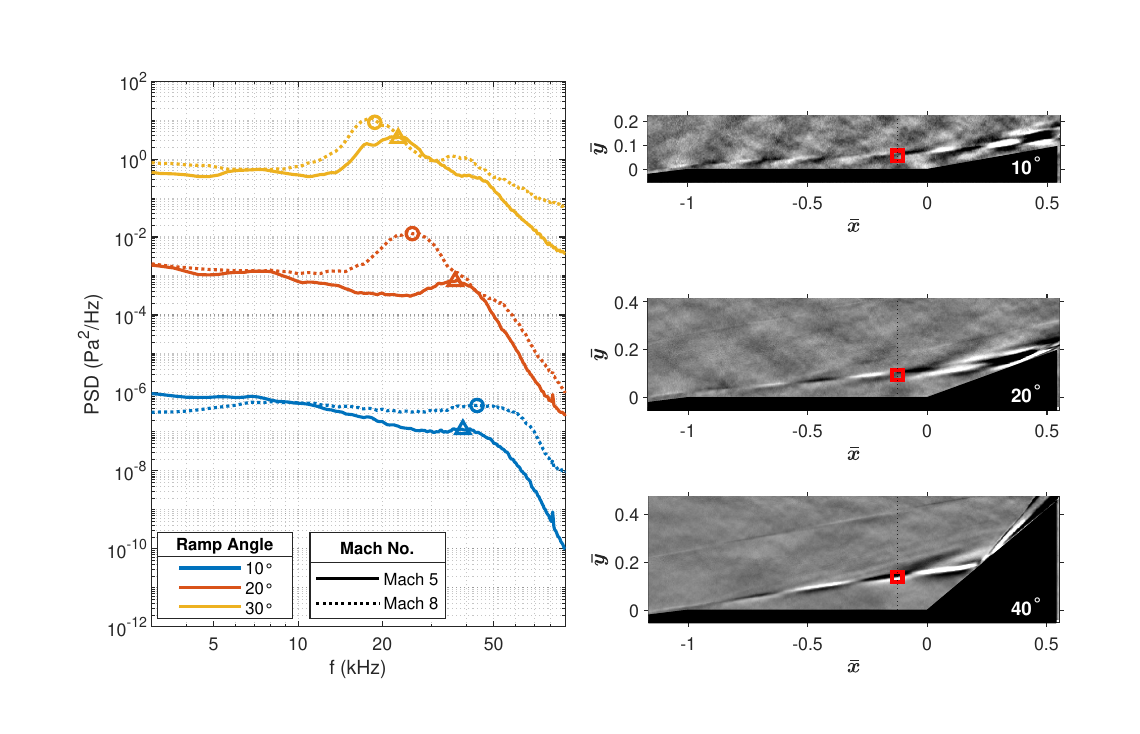}}
		\caption{PSD near ramp corner for different ramp angles and Mach numbers (left) PSDs for $10^{\circ}$ and $20^{\circ}$ ramps have been shifted by $10^{-5}$ and $10^{-2}$, respectively. Schlieren SPOD modes at peak frequency for Mach 8 data (right). $Re \approx 4.0 \times 10^6$ /m}
		\label{fig:PSD_Kul_shearlayer}
	\end{figure}

	\begin{figure} [hbt!]
		\centering
        \subfigure[]
		{\includegraphics[trim={110 260 130 280}, clip, width=0.49\textwidth]{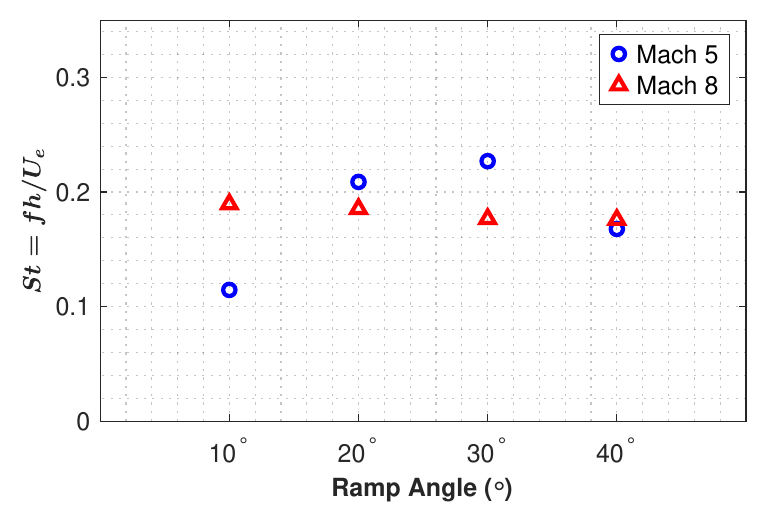} \label{fig:St_vs_rampangle}}
        \subfigure[]
		{\includegraphics[trim={110 260 130 280}, clip, width=0.49\textwidth]{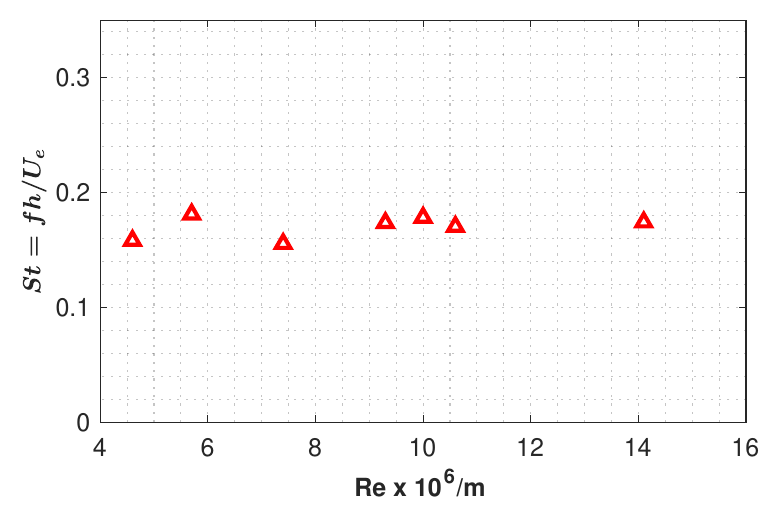} \label{fig:St_vs_Re}}
		\caption{Strouhal number comparison (a) Laminar regime for different ramp angles and Mach numbers, $Re \approx 4.0 \times 10^6$ /m, (b) Mach 8, $30^{\circ}$ ramp with varying Re. Frequencies are taken for the sensor located at $\bar{x} = -0.12$}
		\label{fig:Stplot}
	\end{figure}

     Figure \ref{fig:PSD_Kul_shearlayer} shows the surface-pressure PSD measurements and the corresponding schlieren eigenmodes (shown only for Mach 8) at other ramp angles. The separation bubble extent at these other ramp angles varies from the $30^{\circ}$ case, but the shear-layer unsteadiness is still observed in the PSD and the schlieren. The PSDs correspond to the sensor located at $\bar{x} = -0.13$; sensor data is unavailable for the $40^{\circ}$ case. SPOD modes were picked for the peak frequency observed in the 10-50 kHz range and confirmed the frequency peaks from the pressure sensors as well as provided the frequency for the $40^{\circ}$ case. To verify the shear-layer height versus frequency scaling for these other ramp angles, the height of the shear layer was extracted using the SPOD eigenmodes at $\bar{x} = -0.13$; this location is annotated with red squares in the figure. SPOD modes served better than the mean schlieren images for this purpose especially for the small-ramp angles where the shear layer was obscured by the thick boundary layers. 

    The Strouhal numbers based on the shear-layer height and edge velocity were evaluated for all the ramp angles and both Mach numbers and are presented in figure \ref{fig:St_vs_rampangle}. For the Mach 8 data, the Strouhal number remains nearly constant at 0.18 across all investigated ramp angles. In contrast, the Mach 5 trend in figure \ref{fig:St_vs_rampangle} exhibits greater scatter than that in figure \ref{fig:shearlayer_St_M5M8} due to increased uncertainty in determining the shear-layer height; this is a result of the weaker density gradients associated with the shear-layer structures at Mach 5. While alternative length scales, such as shear-layer thickness, might also collapse the observed frequencies, they could not be estimated with sufficient confidence from the schlieren images. The observed collapse with shear-layer height does not isolate a specific physical origin for the shear-layer unsteadiness, which may be acoustic, entropic, vortical, or a combination of these mechanisms. More importantly, however, the results demonstrate that the separated shear-layer height can effectively predict the characteristic frequency on this geometry, providing a viable basis for reduced-order modeling. 

    To facilitate comparison with two-dimensional laminar SBLI studies, the observed frequencies in figure \ref{fig:PSDcontour_Lam_M8} have been non-dimensionalized using the separation length $L$ and incoming boundary-layer thickness $\delta$ to determine the respective Strouhal numbers, $St_{L}$ and $St_{\delta}$. While the span of the ramp may also be an important length scale in this three-dimensional flow, it is excluded from the present analysis. Because sensor measurements were conducted in the plane of symmetry - where separation region extends from the start of the slice to the middle of the ramp - a separation length of $L \approx 1.25 L_{slice}=117.6$ mm has been adopted. The boundary-layer thickness just upstream of the slice, $\delta$, was estimated from the mean schlieren image as 2.2 mm. The resulting Strouhal numbers, calculated using the cone-edge velocity, have been summarized in Table \ref{tab:St_Lam}.

    \begin{table}[hbt!]
    \caption{\label{tab:St_Lam} Strouhal numbers based on separation length and incoming boundary-layer thickness}
    \centering
    \begin{tabular}{lcccc}
    \hline
    Unsteadiness   & Frequency &  $St_{L}$    &  	$St_{\delta}$   \\\hline
    SU             & 1.4 kHz   &  0.135       & 	0.0025  \\  
	UU  		   & 4.4 kHz   &  0.424 	  & 	0.008 	  \\
	SL  		   & 15-50 kHz &  1.45-4.82   & 	0.03-0.09 	  \\    
    \hline
    \end{tabular}
    \end{table}

    Stability analyses of hypersonic laminar SBLIs have revealed multiple sources of unsteadiness. Instability waves originating in the upstream boundary layer are selectively amplified within the separated shear layer. This amplification is selective to the lower-frequency oblique disturbances \cite{Balakumar2005, Lugrin2021, Dwivedi2022, Paredes2022}. Analysis of Lugrin et al. \cite{Lugrin2021} and Dwivedi et al. \cite{Dwivedi2022} show that these amplified disturbances typically fall within the ranges of $St_{\delta}=0.01-0.09$ and $\lambda_{z}/\delta = 3-10$. Experimentally, these convective instabilities manifest as shear-layer undulations, yielding schlieren SPOD modes similar to those shown in figure \ref{fig:Schlieren_SPOD_Lam_M8} \cite{Benitez2025, Butler2022, Lugrin2022}. 
    
    For steeper ramp angles and higher $Re$, global instabilities can arise spontaneously without upstream disturbances \cite{Sidharth2018, Cao2021, Hao2021, Cao2022}, introducing three-dimensionality and unsteadiness. Hao et al. \cite{Hao2021} suggest that the emergence of global instability is closely linked to secondary separation beneath the primary separation bubble, which is characterized by a localized dip in surface pressure. A similar pressure dip was observed in the present three-dimensional ramp configuration (see figure \ref{fig:MeanP_M5M8}). Furthermore, Cao et al. \cite{Cao2021} identified oscillatory modes at $St_{L}$ values of 0.15 and 0.39 that match unsteady computational modes, while at higher $Re$, multiple oscillatory modes with varying spanwise wavelengths have been reported in the $St_{L}$ range of $0.37-1.28$ \cite{Cao2022}.

    These comparisons suggest that the shear-layer unsteadiness (SL) observed in the present work is likely linked to convective instabilities, while the separation (SU) and the additional unknown (UU) unsteadiness represent global instabilities of the separation region. It must be noted that a comparable Strouhal number does not inherently establish causality and three-dimensional global stability analysis of the computed base flows is necessary to fully resolve the mechanisms driving unsteadiness in this flow field.

    \subsection{Unsteadiness in the transitional regime}
	\label{subsection:UnsteadyTrans}

\begin{figure} [hbt!]
		\centering
		\subfigure[]
		{\includegraphics[trim={108 180 110 179}, clip, width=0.49\textwidth]{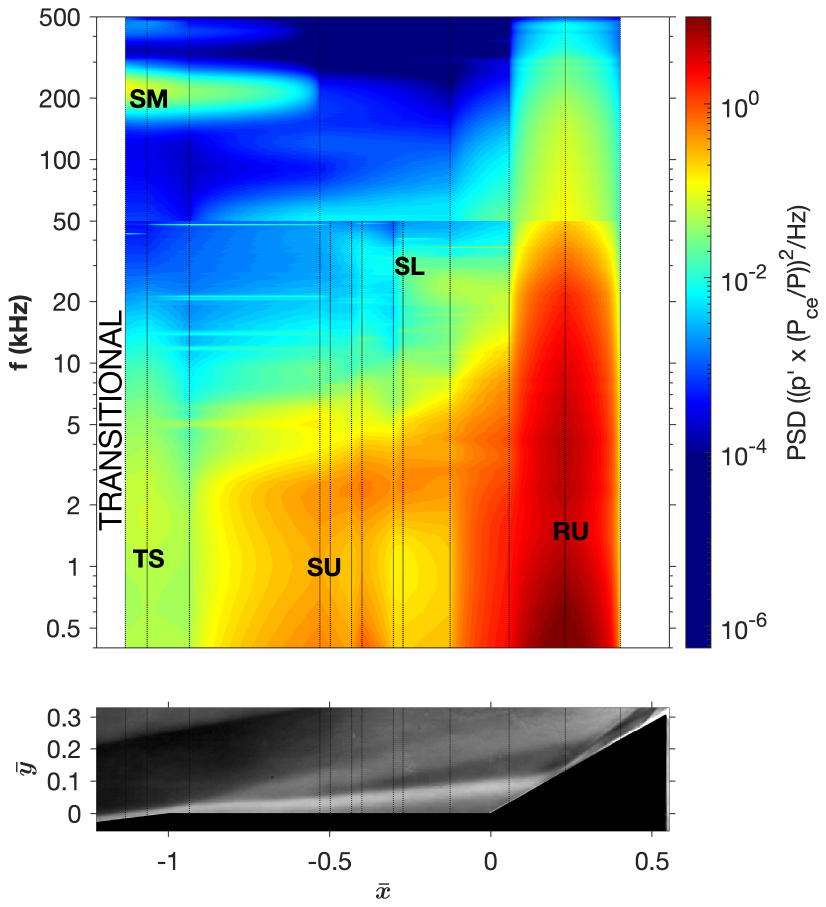} \label{fig:Trans_Re7pt5}}
		\subfigure[]
		{\includegraphics[trim={108 180 110 179}, clip, width=0.49\textwidth]{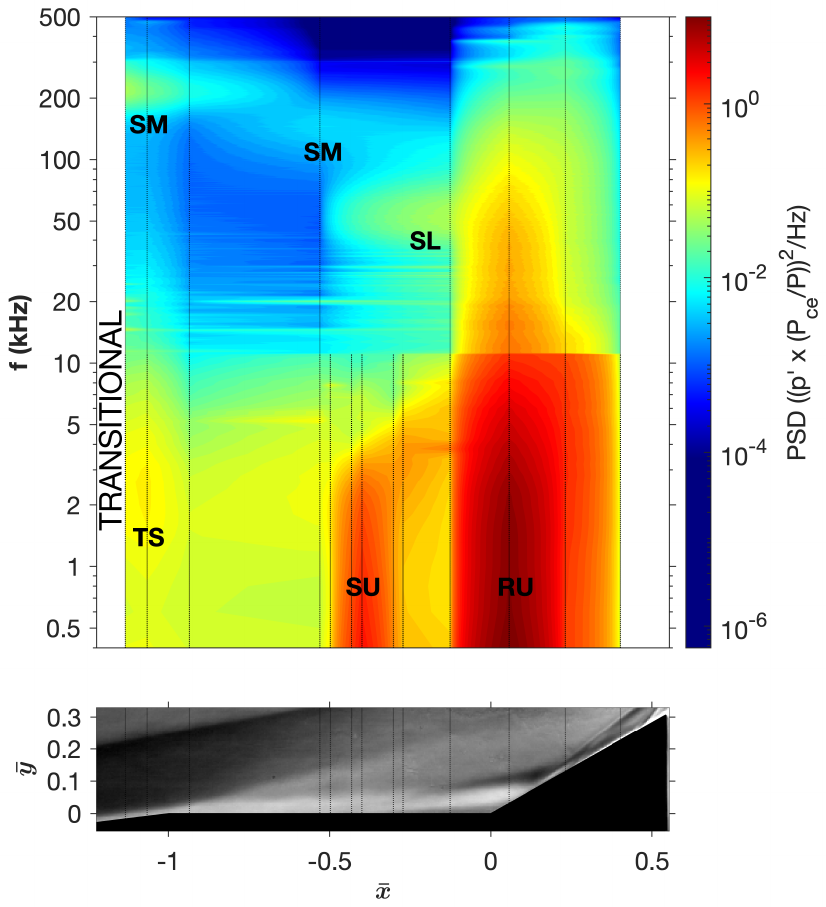} \label{fig:Trans_Re9pt4}}
		\caption{PSD contours from combined Kulite ($0.4-11$ kHz) and PCB ($11-500$ kHz) data for transitional conditions. Mach 8, (a) $Re \approx 7.5\times 10^6$ /m,  (b) $Re \approx 9.4\times 10^6$ /m with corresponding schlieren images. Annotations are discussed in the text.}
		\label{fig:PSDcontour_Trans}
	\end{figure}

\begin{figure} [hbt!]
		\centering                 
		{\includegraphics[trim={0 235 0 240}, clip, width=0.95\textwidth]{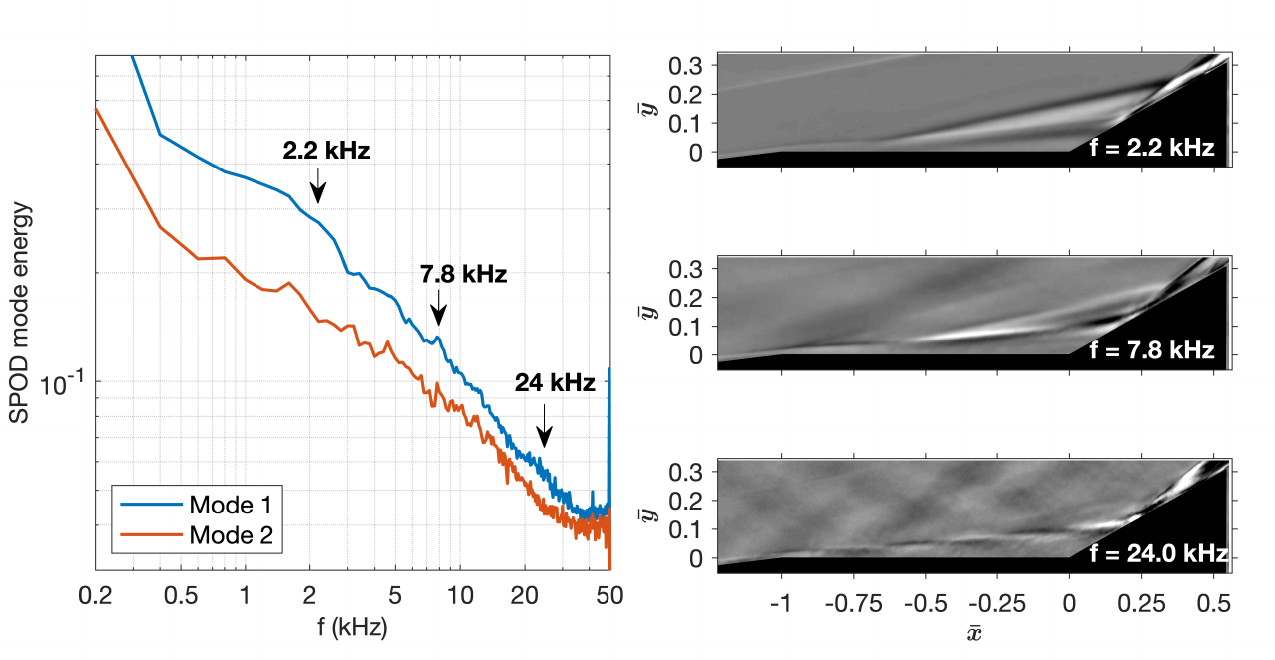}}
		\caption{Schlieren SPOD for Mach 8, $Re = 7.4 \times 10^6$ /m. Eigenmodes at three frequencies are shown.}
		\label{fig:Schlieren_SPOD_Re7pt4_M8}
	\end{figure}

\begin{figure} [hbt!]
		\centering  
        \subfigure[]
        {\includegraphics[trim={108 180 110 179}, clip, width=0.49\textwidth]{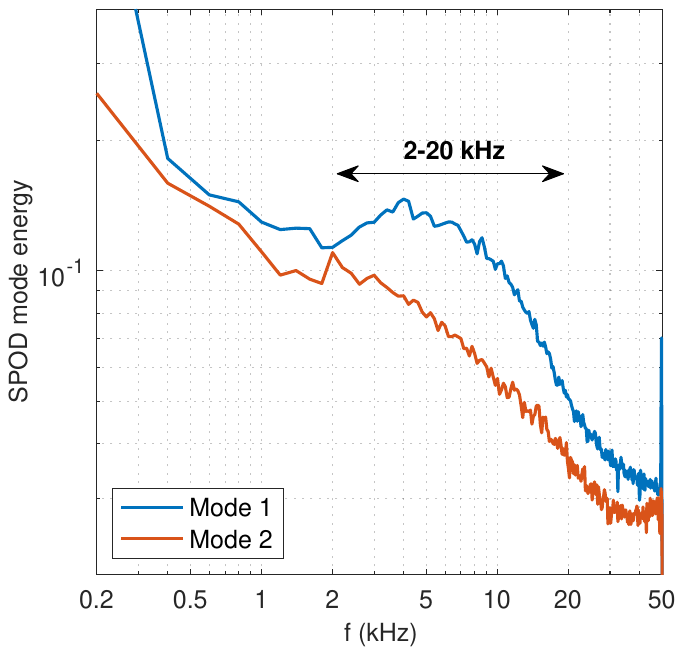}
        \label{fig:Schlieren_SPOD_Re9pt3_M8}}
		        \subfigure[]
        {\includegraphics[trim={108 160 110 160}, clip, width=0.49\textwidth]{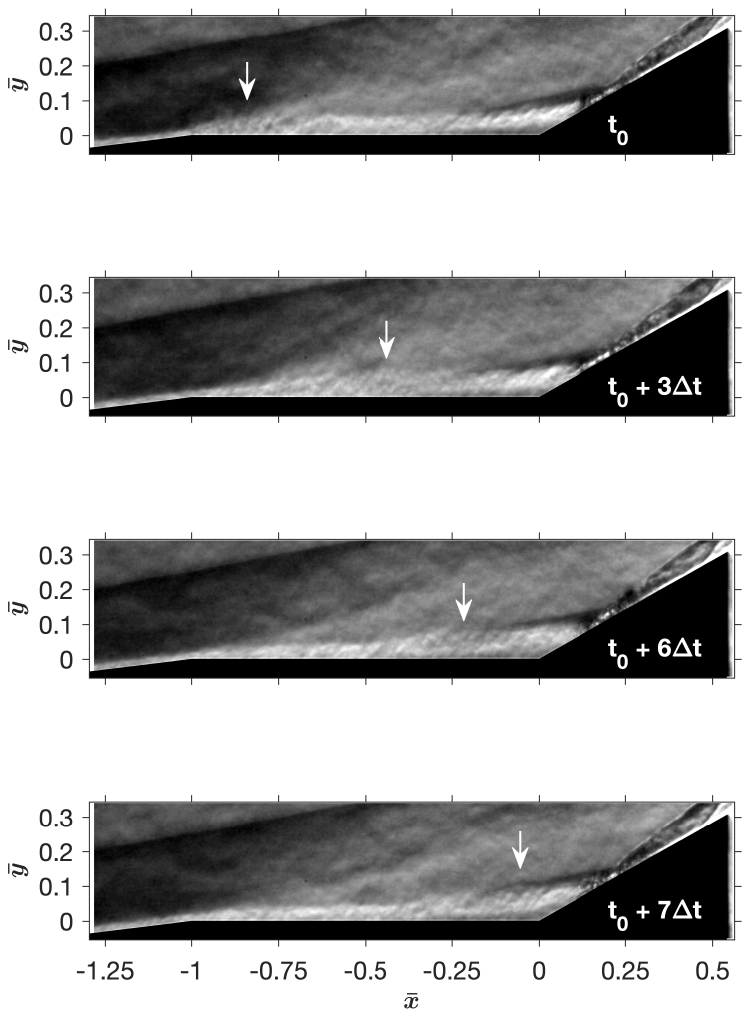}
        \label{fig:Schlieren_Re9pt3_M8_RawSequence}}
		\caption{(a) Schlieren SPOD mode energy for Mach 8, $Re = 9.3 \times 10^6$ /m. (b) Intermittent turbulent spots as observed in individual schlieren images, Mach 8, $Re = 9.3 \times 10^6$ /m.}
		\label{fig:unsteadytransitional}
	\end{figure}

    The transitional regime is characterized by a rapid decrease in separation size and large-scale unsteadiness due to a transition front that breaks down first in the shear layer and then, as $Re$ is increased, in the upstream boundary layer. Surface-pressure PSD contours for two representative cases at $Re \approx 7.5\times 10^6$ /m (left) and $Re \approx 9.4\times 10^6$ /m (right) are shown in figure \ref{fig:PSDcontour_Trans} along with the corresponding mean schlieren images. Where needed, the delineation between the Kulite and PCB data has been moved from 50 kHz (as in figure \ref{fig:PSDcontour_Lam_M8}) to 11 kHz to allow clearer identification of shear layer features that shift in frequency more than for laminar conditions. Corresponding SPOD analysis of the schlieren sequence is shown in figures \ref{fig:Schlieren_SPOD_Re7pt4_M8} and \ref{fig:Schlieren_SPOD_Re9pt3_M8}, respectively, for the two cases.
    
    Since the incoming boundary layer is transitional, strong second-mode instability waves (SM) are registered by the sensors on the cone ($\bar{x} < -1$). These waves are much stronger in the $Re \approx 7.5\times 10^6$ /m case and have begun to break down by $Re \approx 9.4\times 10^6$ /m. The expansion over the slice causes an increase in the boundary-layer thickness and the corresponding second-mode energy decreases and shifts to lower frequencies \citep{Chuvakhov2021, Butler2021ExiF, Pandey2024}.

    For the $Re \approx 7.5\times 10^6$ /m case, very-low-frequency ($<1$ kHz) pressure fluctuations are observed close to the mean separation location of $\bar{x} \approx -0.5$, blending with an approximately 2 kHz unsteadiness which spans a larger streamwise extent. These regions are annotated as SU in figure \ref{fig:Trans_Re7pt5} to highlight this separation unsteadiness. This unsteadiness is also seen as a broad peak in the SPOD spectral energy plot of figure \ref{fig:Schlieren_SPOD_Re7pt4_M8}. A representative eigenfunction at 2.2 kHz shows a large-scale bubble collapse and recharge occurring at this frequency where both the separation shock and the shear layer motion are coupled. In contrast to the $Re \approx 4.0 \times 10^6$ /m case in figure \ref{fig:PSDcontour_Lam_M8}, here the separation point is no longer fixed to the expansion corner, but the large-scale unsteadiness remains prevalent. Using separation length of $L=0.7L_{slice}$ and Prandtl-Meyer derived slice velocity after the expansion (1073 m/s) at this $Re$, $St_L = 0.135$ is estimated which is similar to the value obtained for the SU unsteadiness in the laminar regime.
    
    The sensors within the mean separation region, downstream of SU, also pick up the shear-layer unsteadiness which is annotated as SL in figure \ref{fig:Trans_Re7pt5}. The frequency of approximately 24 kHz near $\bar{x} = -0.12$ is higher than the laminar cases due to the separation location movement off of the expansion corner and into the proximity of the measurement location where the separation region is thinner. This increasing shear-layer frequency scales with the decreasing shear-layer height through the transitional and turbulent regimes  to result in a constant Strouhal number as shown in figure \ref{fig:St_vs_Re} for the Mach 8 data. A minor peak at 24 kHz is seen in the SPOD plot of figure \ref{fig:Schlieren_SPOD_Re7pt4_M8} and the corresponding eigenfunction represents braided shear-layer structures as seen earlier in the laminar datasets. 

    Unlike separation, the fluctuations at reattachment are broadband and extend at least up to 300 kHz which indicates turbulent behavior. However, it appears to be augmented by the unsteadiness originating near separation as particularly strong fluctuations are observed at the lower frequencies as compared to lower $Re$. In correspondence with the upstream movement of the mean reattachment location, the unsteadiness associated with reattachment (annotated as RU) also moves upstream. 
    
    By $Re \approx 9.3\times 10^6$ /m, the transition front is upstream of the slice as seen in the TSP images of figure \ref{fig:TSP_30deg_M8}. The breakdown of second-mode wavepackets results in the intermittent development of turbulent spots. The passage of these turbulent spots into the SBLI region creates associated unsteadiness as demonstrated in figure \ref{fig:Schlieren_Re9pt3_M8_RawSequence}. A few time steps of schlieren images at this $Re$ highlights the passage of the intermittent turbulent spots which are similar to previous observations at Mach 5 \citep{Pandey2023}. These spots retain their coherence as they pass over the expansion corner and through the separation region.  
    
    The Schlieren SPOD spectral energy plot (figure \ref{fig:Schlieren_SPOD_Re9pt3_M8}) highlights the effect of the spot passage. It indicates a peak of energy dominant in the 2-20 kHz range. This frequency range is consistent with the expected frequency of turbulent spot passage during transition. Casper et al. \cite{Casper2019} studied the statistics of these turbulent spots on a slender cone matching the present model upstream of the expansion corner and showed that the effective frequency of the spot passage is in the range of 2-20 kHz. This effect is also captured by the upstream pressure sensor located at $\bar{x} = -1.1$; this is annotated in figure \ref{fig:PSDcontour_Trans} as TS. The SPOD modes look similar to the 7.8 kHz mode shown in figure \ref{fig:Schlieren_SPOD_Re7pt4_M8} and provide a single-frequency view of the broadband turbulent spots.
    
    Finally, the peak shear layer frequency at $\bar{x} = -0.12$ is 52 kHz which is outside the Nyquist limit of the schlieren imaging. However, this fluctuation can be seen, annotated as SL, in the PCB data of figure \ref{fig:Trans_Re9pt4} and follows the scaling in figure \ref{fig:St_vs_Re}. These shear-layer frequencies are distinct from the second-mode instability waves developing in the expanded boundary layer.  

    \subsection{Unsteadiness in the turbulent regime}
	\label{subsection:UnsteadyTurb}

\begin{figure} [hbt!]
		\centering
        \subfigure[]
		{\includegraphics[trim={108 180 110 179}, clip, width=0.49\textwidth]{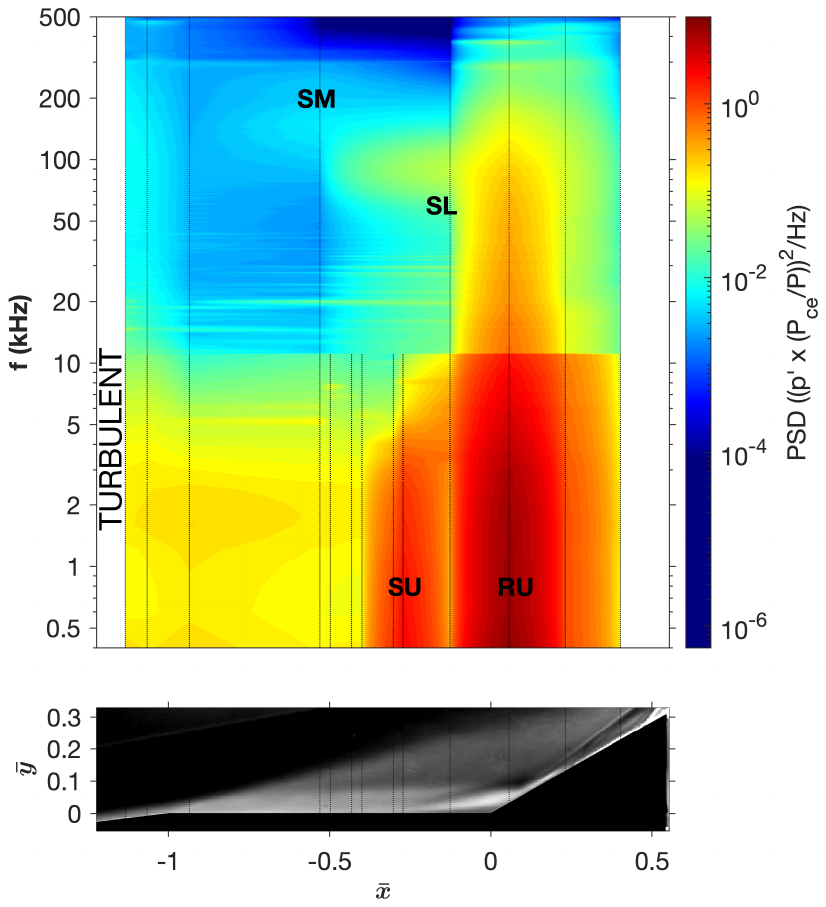} \label{fig:PSD_contour_M8_Turb_16pt5}}
        \subfigure[]
        {\includegraphics[trim={130 220 150 230}, clip, width=0.49\textwidth]{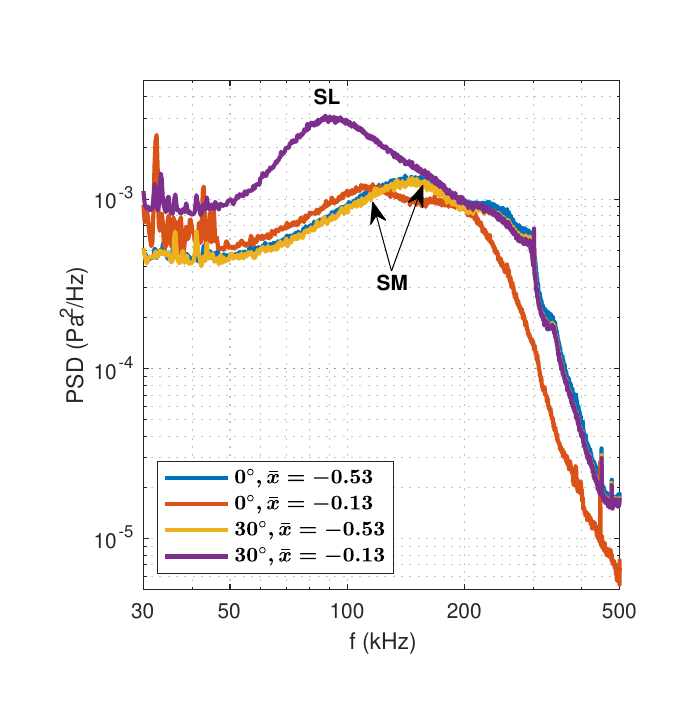} \label{fig:PSD_M8_Turb_0vs30}}
		\caption{(a) PSD contours from combined Kulite ($0.4-11$ kHz) and PCB ($11-500$ kHz) data for turbulent conditions at Mach 8, $Re \approx 16.5\times 10^6$ /m with corresponding schlieren images. Annotations are discussed in the text. (b) Comparison of spectra on the slice, with and without the ramp.}
		\label{fig:PSDcontour_Turb_M8}
	\end{figure}

    As discussed earlier, the expansion corner induces strong relaminarization effects at Mach 8 that cause an incoming turbulent boundary layer to become transitional on the slice. The PSD contours for $Re \approx 16.5 \times 10^6$ /m are shown in figure \ref{fig:PSDcontour_Turb_M8}. Broadly similar features as in figure \ref{fig:Trans_Re9pt4} can be observed although with a few minor differences. Also shown is a comparison of the PSD for the PCB sensors for the cases with and without the ramp through use of the sensors located upstream and downstream of the 30-deg separation location. 
	
    First, the second-mode waves are absent in the incoming boundary layer as expected due to the broadband behavior typical of turbulent boundary layers. Due to relaminarization effects, second-mode waves reappear on the slice and are annotated with SM. This new second-mode peak is at about 100-150 kHz and reflects the formation of a new instability wave in the thickened boundary layer post-expansion. 
    
    Similar to figure \ref{fig:Trans_Re9pt4}, dominant separation (SU) and reattachment unsteadiness (RU) are observed in the low-frequency regime. However, the separation unsteadiness is further downstream which is in agreement with the downstream movement of the separation location as seen in section \ref{subsection:ScaleofSep}. Such low-frequency unsteadiness near separation is a well-documented phenomenon in two-dimensional turbulent SBLIs \citep{Clemens2014, Gaitonde2023ARFM} where these occur at a nominal Strouhal number of 0.03 based on separation length and edge velocity. Using Prandtl-Meyer estimated velocity after the expansion (1060 m/s) and experimentally-observed separation lengths of $\bar{x} = 0.2 - 0.4$, this Strouhal number corresponds to separation-shock frequencies of 850--1690 Hz which are within the range of SU observed in figure \ref{fig:PSD_contour_M8_Turb_16pt5}. Therefore, the low-frequency separation motion of the present noncanonical geometry under turbulent conditions appears well predicted by a two-dimensional compression ramp.
    
    Finally, shear-layer unsteadiness (SL) is captured at approximately 90 kHz by the sensor located at $\bar{x} = -0.13$ which is within the separation bubble. The frequency is higher compared to those noted in figures \ref{fig:PSDcontour_Lam_M8} and \ref{fig:PSDcontour_Trans} given that the mean-separation location has moved closer to this measurement location at this higher Re. This corresponds to a lower shear layer height and therefore a higher SL frequency.

    The fluctuations induced by the ramp are directly compared with the no-ramp case in figure \ref{fig:PSD_M8_Turb_0vs30} to evaluate the differences in the evolution of turbulence. The sensor located at $\bar{x} = -0.53$, which is upstream of separation, demonstrates very similar behavior across the two experiments: a broad peak due to incipient second-mode waves at 160 kHz and a sharp roll-off at higher frequencies ($>300$ kHz) due to decay of near-wall turbulence. For the expansion-only case, the boundary layer continues to evolve and the downstream sensor located at $\bar{x} = -0.13$ demonstrates a further decay in high-frequency fluctuations and the second-mode frequency shifts to a lower frequency of about 120 kHz. In contrast for the separated scenario, the fluctuation levels get amplified by the separation shock for $<150$ kHz range. A dominant peak appears at about 90 kHz which is attributed to the shear-layer unsteadiness as discussed earlier.

	\subsection{Variations with Mach number}
	\label{subsection:UnsteadyCompilation}

\begin{figure}
		\centering
		\subfigure[]
		{\includegraphics[trim={90 220 80 235}, clip, width=0.49\textwidth]{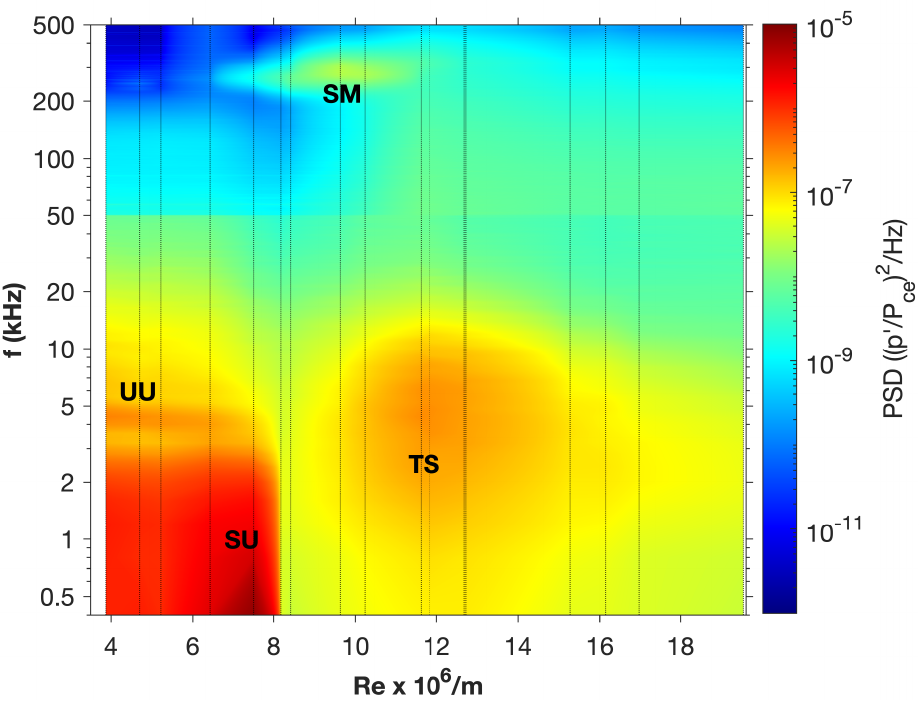} \label{fig:PSDcontourM5_xminus88}}
		\subfigure[]
		{\includegraphics[trim={90 220 80 235}, clip, width=0.49\textwidth]{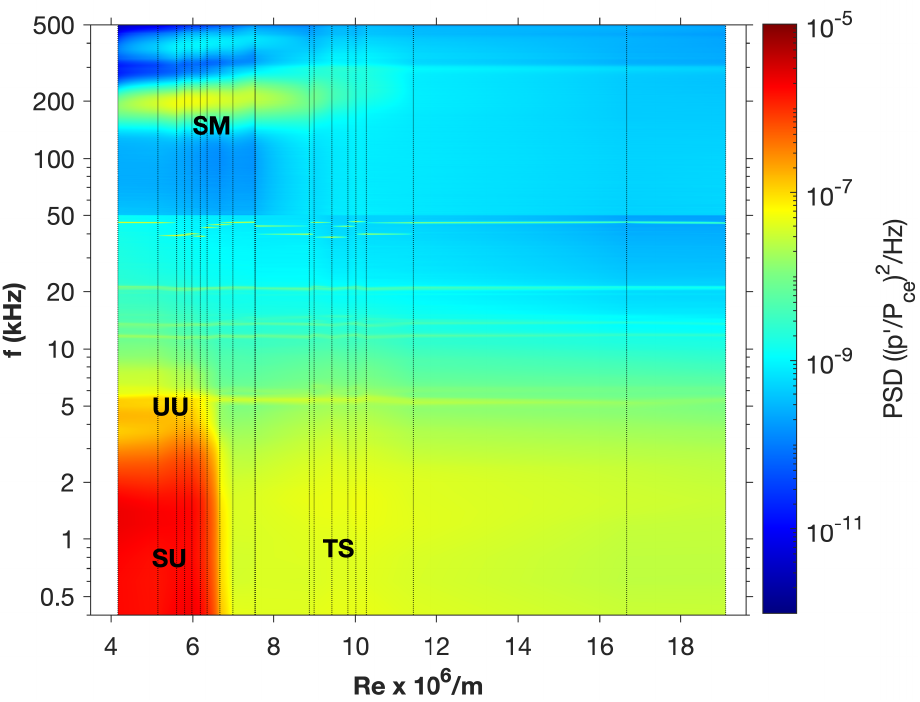} \label{fig:PSDcontourM8_xminus88}}
        \subfigure[]
        {\includegraphics[trim={90 220 80 235}, clip, width=0.49\textwidth]{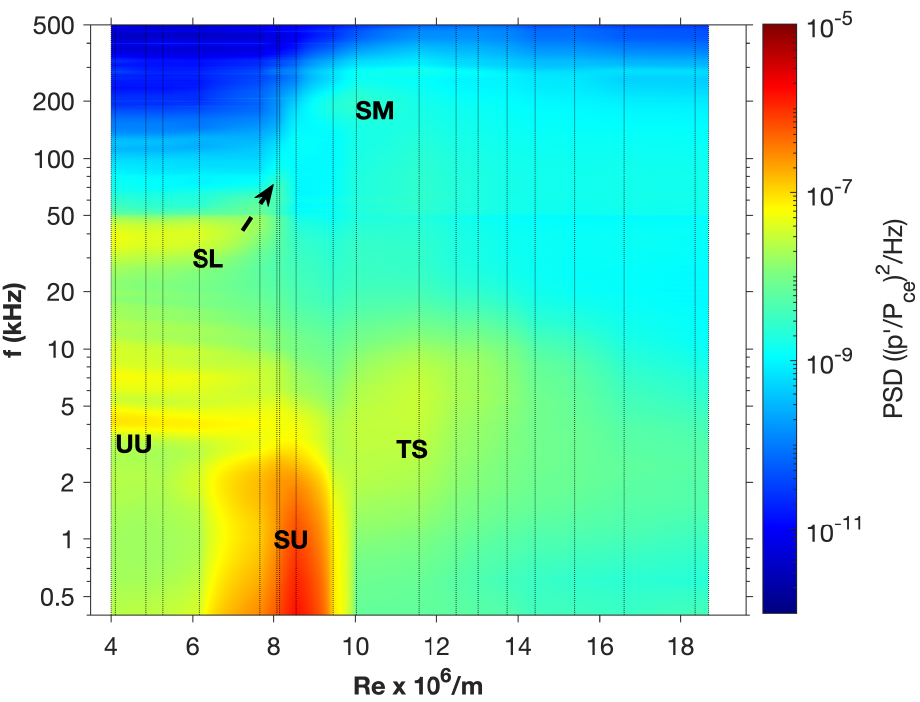} \label{fig:PSDcontourM5_xminus50}}
		\subfigure[]
		{\includegraphics[trim={90 220 80 235}, clip, width=0.49\textwidth]{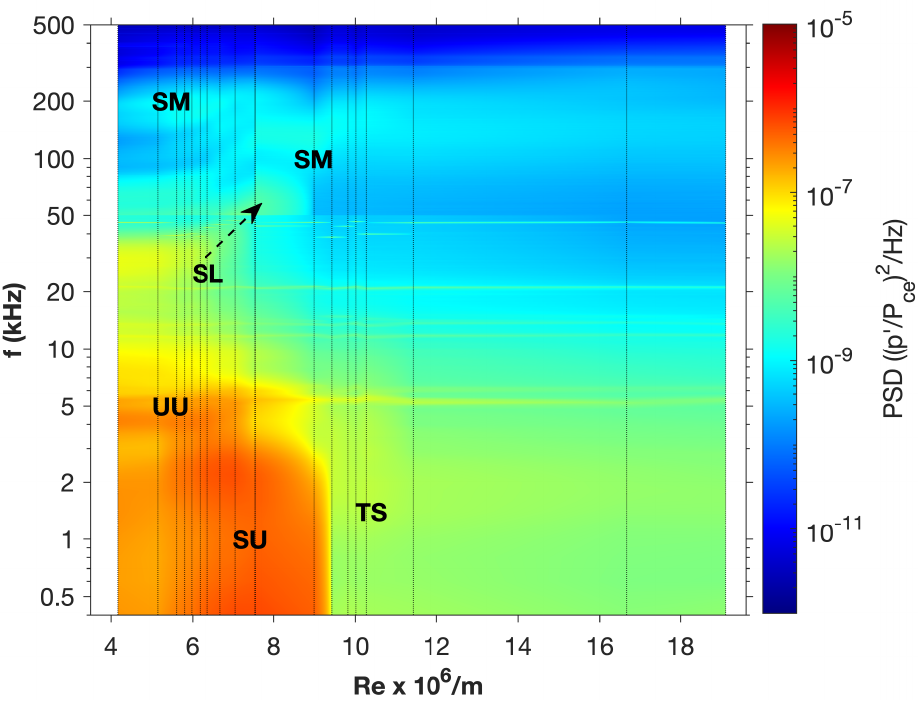} \label{fig:PSDcontourM8_xminus50}}
        \subfigure[]
        {\includegraphics[trim={90 220 80 235}, clip, width=0.49\textwidth]{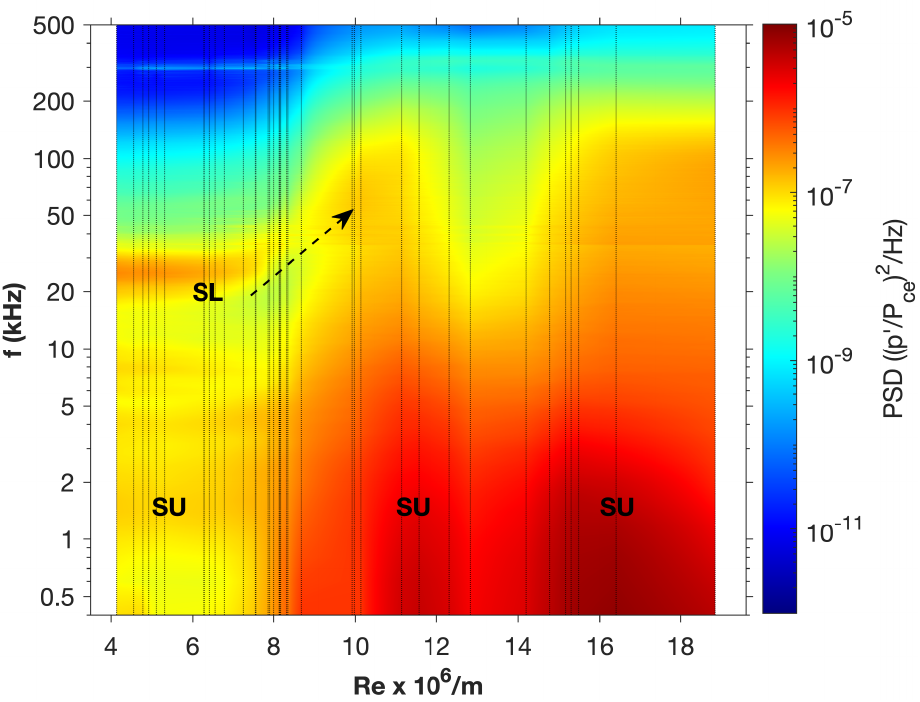} \label{fig:PSDcontourM5_xminus11}}
		\subfigure[]
		{\includegraphics[trim={90 220 80 235}, clip, width=0.49\textwidth]{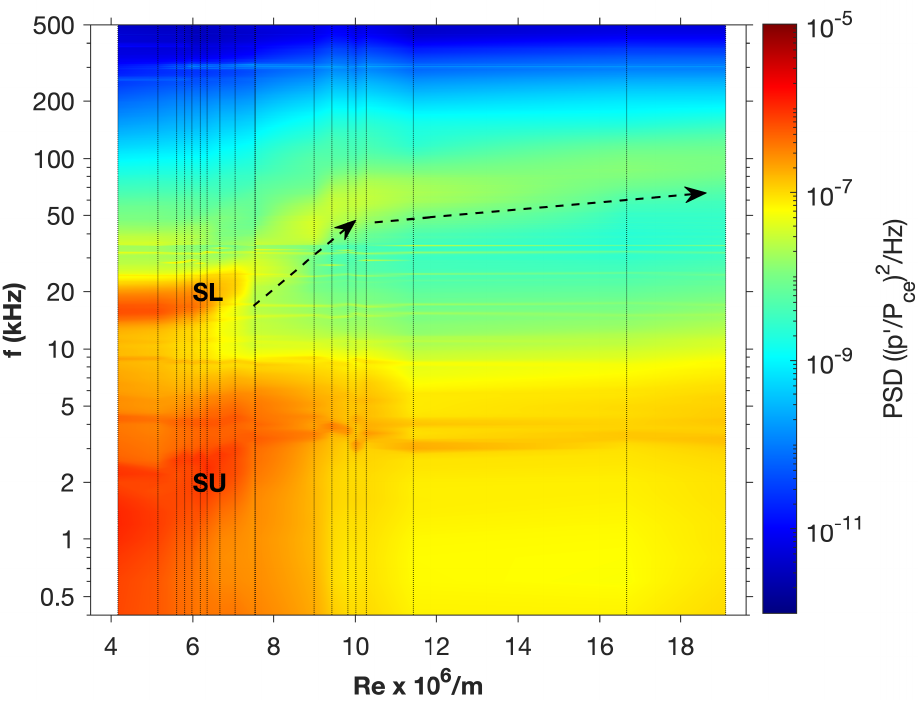} \label{fig:PSDcontourM8_xminus11}}
		\caption{PSD contours at Mach 5 (left) and Mach 8 (right) from combined Kulite ($0.4-50$ kHz) and PCB ($50-500$ kHz) data at $\bar{x} = -0.88$ (a) and (b), $-0.53$ (c) and (d), and $-0.11$ (e) and (f). Dotted lines represent the $Re$ tested. Annotations are discussed in the text.}
		\label{fig:PSDcontour_Re_Mvar}
	\end{figure}

    The previous sub-sections provided details on the types of unsteadiness present at representative $Re$ cases for laminar, transitional and turbulent incoming flow to the expansion corner at Mach 8. In this section, select sensors on the slice have been used to present the evolution of fluctuations at fixed spatial locations as the SBLI scale changes with increasing $Re$ and Mach number. Figure \ref{fig:PSDcontour_Re_Mvar} presents PSD contours at Mach 5 (left) and Mach 8 (right) at $\bar{x} = -0.88, -0.53, -0.11$. The evolution of the pressure fluctuations upstream of the slice and on the ramp do not contain any new modes of fluctuations and their description has been omitted for brevity. Dotted lines indicate the $Re$ of the various HWT runs; caution should be exercised in interpolated data further away from those lines. The PSD contours have been normalized using the cone-edge pressure calculated using the tunnel freestream conditions and Taylor-Maccoll equations. This is in contrast to the additional normalization by mean pressure on earlier spectrograms and therefore the contour levels are different. 

    The discussions from the previous subsections help elucidate the features in the Mach 8 dataset. As seen in figure \ref{fig:PSDcontourM8_xminus88} for location $\bar{x} = -0.88$, the low $Re$ cases ($Re < 6.7 \times 10^6$ /m) show the low-frequency separation unsteadiness (SU) along with the 4.4 kHz unsteady motion (UU) and second-mode (SM) fluctuations as described in \ref{subsection:UnsteadyLam}. For a further increase in $Re$ ($Re > 6.7 \times 10^6$ /m), these low-frequency fluctuations disappear abruptly as the shock and separation locations move downstream of this sensor and only the second-mode waves remain in the incoming attached, transitional boundary layer. As $Re$ increases further, these instability waves breakdown and intermittent passage of boundary-layer transitional structures or turbulent spots (TS) are observed faintly at $Re \approx 10 \times 10^6$ /m. No other features can be discerned at higher $Re$ at this location.
    
    In comparison, the Mach 5 data set, figure \ref{fig:PSDcontourM5_xminus88}, demonstrates broadly similar behavior at this location but the relative strength of the features and $Re$ over which they are prominent differs from Mach 8. The low-frequency separation unsteadiness (SU) is present but extends to a higher $Re$ before the shock and separation locations move downstream. Second-mode waves are observed at a higher frequency than at Mach 8 and also initiate at a higher $Re$. The impact of the turbulent spots (TS) that was faint in figure \ref{fig:PSDcontourM8_xminus88} is much stronger here. These differences are likely because Mach 5 transition occurs due to a mixture of first and second mode waves \citep{Casper2016, Hader2021} whereas Mach 8 transition is driven by pure second-mode instabilities. Finally, the unknown unsteadiness is more visible at this lower Mach number.
    
    
    In the middle of the slice, the sensor located at $\bar{x} = -0.53$ measures the fluctuations induced by the shear layer. As seen in figure \ref{fig:PSDcontourM8_xminus50}, the Mach 8 shear layer fluctuations (SL) peak in the laminar regime at $\approx29$ kHz and stay constant before increasing as annotated with a black dashed arrow. The 17 kHz peak frequency of figure \ref{fig:PSDcontour_Lam_M8} occurs downstream of the $\bar{x} = -0.53$ sensor examined here and is therefore absent. In addition to the shear layer unsteadiness, low-frequency separation unsteadiness is also present over much of the laminar and transitional regimes.  A further increase in $Re$ results in the abrupt disappearance of these fluctuations as the separation sweeps downstream of this sensor. This occurs at a higher $Re$ than in figure \ref{fig:PSDcontourM8_xminus88} because the sensor location further downstream in figure \ref{fig:PSDcontourM8_xminus50} detects fluctuations once the shock and separation locations begin to move downstream.
    
    In comparison, the Mach 5 contours at $\bar{x} = -0.53$ shown in figure \ref{fig:PSDcontourM5_xminus50} exhibit peak shear-layer fluctuation (SL) frequencies of $\approx$35 kHz which is higher than for the Mach 8 case because of the correspondingly smaller separation region. The 4.4 kHz unsteadiness (UU) is also present. In addition, the low-frequency separation unsteadiness (SU) before the flow has become fully turbulent remains the dominant feature at both Mach numbers. The laminar regime persists at the lowest $Re$ before the separation unsteadiness becomes significant at this location. The unsteady effects due to turbulent spot passage are observable at this location as well.

    Figure \ref{fig:PSDcontourM8_xminus11} shows the Mach 8 contour at $\bar{x} = -0.11$; notable is the extended range of frequencies measured for the shear-layer unsteadiness (SL). These begin at a constant value of $\approx17$ kHz in the laminar regime before increasing in the transitional regime. In line with the mean SBLI changes discussed in the context of figures \ref{fig:fig_Schlieren_M8} and \ref{fig:TSP_30deg_M8}, the frequency increases rapidly in the transitional regime before slowing down as the upstream boundary layer becomes turbulent. This change is annotated with black dashed arrow in figure \ref{fig:PSDcontourM8_xminus11}. In addition, separation unsteadiness (SU) in the laminar regime can be observed due to upstream separation propagating fluctuations through the separation region and over this sensor location. 
    
    In comparison, the Mach 5 contour in figure \ref{fig:PSDcontourM5_xminus11} shows a much different distribution because the separation location first sweeps downstream and then back upstream as the $Re$ increases. As a result, two lobes of the separation unsteadiness with a gap in between can be observed in figure \ref{fig:PSDcontourM5_xminus11}. This different behavior is expected because as discussed previously, relaminarization effects are much stronger at Mach 8. As a result, only Mach 5 exhibits fully turbulent separation behavior at high $Re$. The smaller separation region can cross over the sensor at this location, leading to stronger fluctuations than at Mach 8.

    Overall, similar features are found at Mach 5 as in the Mach 8 data more heavily discussed in the present publication. The frequencies of these features shift due to differences in the boundary layer and shear layer as well as flow velocity. Only Mach 5 reaches a turbulent shock boundary layer interaction due to the weaker relaminarization across the expansion corner. The presence of the upstream expansion on this noncanonical geometry plays a substantial role in the behavior of the SBLI at both Mach numbers.

	\section{Conclusion}
	
    The hypersonic flowfield around a three-dimensional expansion-compression geometry was experimentally studied in the Sandia Hypersonic Wind Tunnel using a suite of diagnostic techniques. Experiments were conducted at two Mach numbers and a range of Reynolds numbers to characterize the laminar, transitional, and turbulent SBLI forming on this non-canonical geometry. 
    
    To characterize the oncoming flow that generates the SBLI, mean flow and unsteady characteristics over the expansion-only ($0^{\circ}$ ramp) geometry were first studied. As the flow from the conical frustrum passes over the expansion corner, the boundary layer thickens considerably, surface streamlines curve away from the streamwise direction, and boundary-layer fluctuations decay at turbulent conditions. At transitional conditions, the second-mode instability waves weaken. These effects are weaker at the lower Mach number because the strength of the expansion scales with the upstream Mach number and the recovery length is shorter. 
    
    A $30^{\circ}$ ramp was introduced onto the slice and the effect of Reynolds number variation on the time-averaged SBLI flow was studied. In the laminar regime, separation was observed to be locked at the expansion corner, encompassing the entire slice until reattachment in the middle of the ramp. Some differences in separation extent were observed with Mach number, where the separation boundary spilled over the sides of the slice. As $Re$ increased and the flow transitioned, the SBLI region shrank rapidly at both Mach numbers. Peak non-dimensional heating rates were observed in the laminar and the early transitional regimes and were observed downstream of reattachment location. In high $Re$ cases, the Mach 5 data demonstrated a sharp increase in the heating near separation which is a signature of a turbulent separation. On the other hand, the SBLI size continued to decrease at Mach 8 even in the highest $Re$ cases and the separation region showed a characteristic dip in heating which is a signature of transitional separation. These results indicated that the stronger relaminarization effects at Mach 8 precluded the emergence of a turbulent SBLI on this geometry even in the nominally turbulent $Re$ range tested here.
    
    Unsteady effects were studied across the different flow regimes using surface pressure measurements and schlieren visualizations. In the laminar regime, low-frequency separation unsteadiness was observed below about 4 kHz with coupled motion of the separation shock and shear layer. This occurred with the separation shock foot pinned to the expansion corner. Another fluctuation at 4.4 kHz occured under low $Re$ conditions whose origin is uncertain at this time. A shear-layer flapping unsteadiness was also observed with a frequency continuously evolving in the streamwise direction and a peak amplitude near the slice-ramp corner. This unsteadiness was observed with ramps of different deflection angle, at both Mach numbers, and across a large range of Reynolds numbers. Peak frequencies were observed at approximately 17 kHz near reattachment but could reach beyond 50 kHz at other points within the separation region. The shear-layer frequencies across this large dataset could be collapsed using shear-layer height as the length scale and edge velocity as the velocity scale. Finally, second-mode instability waves were transmitted along the separated shear layer onto the reattached boundary layer on the ramp. 
    
    In the early transitional regime, the separation shock moves off the expansion corner and is located on the slice itself. This shrinks the size of the separation region and, the low-frequency unsteadiness below approximately 4 kHz is again observed with large-scale coupled movement of the separation shock and the shear layer. As the upstream boundary layer breaks down into turbulent spots, their passage into the SBLI region creates large unsteadiness of the separated flow extending from 2-20 kHz. The influence of turbulent spots on the spectral content was found to be stronger at Mach 5 than Mach 8. The frequency of shear-layer unsteadiness increases in the transitional regime and second-mode waves continue to be observed passing through the separation region transmitted along the shear layer.
    
    Finally, the high-$Re$ regime at Mach 8 retained a transitional behavior due to the strong relaminarization and followed the unsteady mechanisms observed at lower $Re$. The separation shock was located further downstream even though the flow cannot be considered fully turbulent due to persistent relaminarization effects. The Mach 5 condition, on the other hand, exhibited a turbulent SBLI due to the weaker relaminarization effects. After accounting for the changes in the mean-flow variation with $Re$, the Mach 5 dataset demonstrated similar types of unsteadiness in comparison to Mach 8.

    Many of the SBLI features long known from the extensive literature of two-dimensional compression ramps are found here. This includes low-frequency breathing motions of the separation bubble that create shock unsteadiness, dramatic pressure rises within the separation region, and increases in mean heat flux at reattachment.  In the present case, however, the strong relaminarization as the oncoming flow passes over the expansion corner adds unique effects to this noncanonical geometry.  Such considerations are important to developing predictive tools for practical flight applications.
       	
\section*{Acknowledgments}
    Over the course of this study, Charley Downing, Thomas Grasser, John Henfling, Ashley Saltzman, Melissa Soehnel and Seth Spitzer helped with model preparation, HWT testing, and data acquisition. Helpful conversations with Lawrence Dechant, Daning Huang, Aravinth Sadogapan, Adam Jirasek and Jurgen Seidel are gratefully acknowledged.

    \section*{Funding Sources}
Sandia National Laboratories is a multi-mission laboratory managed and operated by National Technology and Engineering Solutions of Sandia, LLC., a wholly owned subsidiary of Honeywell International, Inc., for the U.S. Department of Energy’s National Nuclear Security Administration under contract DE-NA0003525. The views expressed in the article do not necessarily represent the views of the U.S. Department of Energy or the United States Government.

\bibliography{references}

\end{document}